\documentclass[a4paper,10pt]{revtex4}
\usepackage{graphicx}  
\usepackage{amsmath}   
\usepackage{amssymb}   
\usepackage{bm} 
\usepackage{dcolumn}
\usepackage{color}
\usepackage{mathrsfs}
\usepackage{amsfonts}
\usepackage{varioref}
\usepackage{mathrsfs}
\usepackage{graphicx}
\usepackage{latexsym}
\usepackage{amsmath}
\usepackage{amssymb}
\usepackage{textcomp}
\usepackage{amsbsy}
\usepackage{graphics}
\usepackage{epstopdf}
\usepackage{color}
\usepackage[caption=false]{subfig}

\RequirePackage[colorlinks,citecolor=blue,urlcolor=magenta,linkcolor=blue]{hyperref}
\input epsf

\allowdisplaybreaks[4]

\begin{document}

\tolerance=5000

\title{Holographic realization from inflation to reheating in generalized entropic cosmology}

\author{Sergei~D.~Odintsov$^{1,2}$\,\thanks{odintsov@ieec.uab.es},
Simone~D'Onofrio$^{2}$\,\thanks{donofrio@ice.csic.es},
Tanmoy~Paul$^{3,4}$\,\thanks{pul.tnmy9@gmail.com}} \affiliation{
$^{1)}$ ICREA, Passeig Luis Companys, 23, 08010 Barcelona, Spain\\
$^{2)}$ Institute of Space Sciences (ICE, CSIC) C. Can Magrans s/n, 08193 Barcelona, Spain\\
$^{3)}$ National Institute of Technology Jamshedpur, Department of Physics, Jamshedpur - 831 014, India\\
$^{4)}$ Labaratory for Theoretical Cosmology, International Centre of Gravity and Cosmos,
Tomsk State University of Control Systems and Radioelectronics (TUSUR), 634050 Tomsk, Russia.}


\tolerance=5000

\begin{abstract}
The growing cosmological interest of different entropy functions (like the Tsallis entropy, the R\'{e}nyi entropy, the Barrow entropy, the Sharma-Mittal entropy, the Kaniadakis entropy and the Loop Quantum gravity entropy) naturally raises an important question: "Does there exist a generalized entropy that can bring all the known entropies proposed so far within a single umbrella?" In spirit of this, recently a four parameter generalized entropy has been formulated that reduces to different known entropies for suitable limits of the parameters. Based on such four parameter generalized entropy (symbolized by $S_\mathrm{g}$), in the present paper, we examine the universe's evolution during its early phase, particularly from inflation to reheating, in the context of entropic cosmology where the entropic energy density acts as the inflaton. It turns out that the entropic energy successfully drives an early inflationary phase with a graceful exit, and moreover, the theoretical expectations of the observable indices get consistent with the recent Planck data for suitable ranges of the entropic parameters. After the inflation ends, the universe enters to a reheating stage when the entropic energy decays to relativistic particles with a certain decay rate. Actually the presence of the entropic parameters in the $S_\mathrm{g}$ ensures a continuous evolution of the Hubble parameter from a quasi de-Sitter phase during the inflation to a power law phase during the reheating stage dominated by a constant EoS parameter. Consequently we investigate the reheating phenomenology, and scan the entropic parameters from both the inflation and reheating requirements. We further address the possibility of instantaneous reheating in the present context of generalized entropy.
\end{abstract}

\maketitle

\section{Introduction}

The hallmark of the Bekenstein-Hawking entropy of a black hole is that it connects the laws of thermodynamics with spacetime gravity \cite{Bekenstein:1973ur,Hawking:1975vcx} (also see \cite{Bardeen:1973gs,Jacobson:1995ab,Wald:1999vt,Cai:2005ra,Padmanabhan:2009vy,Verlinde:2010hp,Padmanabhan:2003gd,Paranjape:2006ca,Akbar:2006kj}). In this arena of gravity-thermodynamics, entropy plays a significant role and can be regarded as one of the most fundamental physical quantities. However unlike to classical thermodynamics where the entropy is proportional to volume of the system under consideration, the black hole entropy turns out to be proportional to the horizon area of the black hole. Based on this distinctive nature, recent literatures proposed different entropy functions, like the Tsallis entropy and the R\'{e}nyi entropy depending on the non-additive statistics \cite{Tsallis:1987eu,Renyi}. The Barrow entropy has been proposed in \cite{Barrow:2020tzx}, which encodes the fractal structure of the black hole that may originate from quantum gravity effects. Some other entropy functions are Sharma-Mittal entropy (a possible combination of the Tsallis entropy and the R\'{e}nyi entropy) \cite{SayahianJahromi:2018irq}, Kaniadakis entropy \cite{Kaniadakis:2005zk,Drepanou:2021jiv} and the entropy in the context of Loop Quantum gravity \cite{Majhi:2017zao,Liu:2021dvj} etc. All of these entropies prove to be function of the Bekenstein-Hawking entropy variable ($S$) and share the following properties: (a) monotonically increase with respect to the Bekenstein-Hawking entropy and (b) vanishes in the limit $S \rightarrow 0$. Owing to these common properties, one may naturally ask the question like -- ``does there exist a generalized entropy that can bring all the known entropies proposed so far within a single umbrella?'' Here it deserves mentioning that the microscopic origin of horizon entropy is still questionable (in this regard, one may see some recent development in \cite{Nojiri:2023ikl}). In absence of a proper microscopic origin, the horizon entropy may be thought as due to some energy flux of the matter fields from inside to outside of the horizon. Since the cosmic horizon (generally given by $r_\mathrm{H} = 1/H$ which is a null surface in a spatially flat Friedmann-Lema\^{i}tre-Robertson-Walker spacetime) divides the observable universe from the unobservable one, such energy flux of the matter fields can be regarded as some kind of information loss which in turn gives rise to the entropy of horizon \cite{Cai:2005ra,Akbar:2006kj,Sanchez:2022xfh}. From a different perspective, the authors of \cite{Komatsu:2013qia,Elizalde:2019jmh} showed that for entropy $\propto r_\mathrm{H}^3$, the energy flux of matter fields from inside to outside of the horizon, in particular, the entropic force (defined by the derivative of the energy flux with respect to the horizon radius) can be associated with some bulk viscosity which may act as the source for generating entropy in the homogeneous and isotropic universe.

Consequently the thermodynamics of horizon is extended to the field of cosmology, in which case, the cosmological horizon is generally defined by the inverse of the Hubble parameter of the universe. Being directly related to the entropic cosmology, the holographic cosmology, initiated by Witten and Susskind \cite{Witten:1998qj,Susskind:1998dq,Fischler:1998st}, earned a lot of attention now a days. Some of our authors recently showed that entropic cosmology of various entropy functions can be equivalently mapped to generalized holographic scenario with suitable holographic cut-offs which depend on either the particle horizon and its derivatives or the future horizon and its derivatives \cite{Nojiri:2021iko,Nojiri:2021jxf}. One of the significant contributions of holographic cosmology corresponding to the aforementioned entropies is the explanation of the dark energy era of our universe, namely the holographic dark energy (HDE) \cite{Li:2004rb,Li:2011sd,Wang:2016och,Nojiri:2005pu,Enqvist:2004xv,Gong:2004cb,Gao:2007ep,Li:2009bn,Zhang:2009un,Lu:2009iv,Sheykhi:2023aqa,Komatsu:2016vof,Nojiri:2019skr,Barrow:2020kug,Bhardwaj:2021chg,Chakraborty:2020jsq,Sarkar:2021izd,Sheykhi:2022jqq}. Note that most general HDE is introduced in \cite{Nojiri:2005pu}, it was explicitly demonstrated in \cite{Nojiri:2021iko} that any HDE (including some scalar field dark energy models) is just particular example of generalised HDE with suitale cut-off. In the context of HDE, we do not need to put some scalar field by hand in Lagrangian, actually the holographic energy density with a suitable cut-off does the job. Beside the dark energy era, the holographic cosmology also keeps the signature in describing the inflationary scenario during the early universe \cite{Nojiri:2022aof,Nojiri:2019kkp,Paul:2019hys,Elizalde:2019jmh,Oliveros:2019rnq,Chakraborty:2020tge,Cai:2010zw,Cai:2010kp,gargee,Mohammadi:2021gvf,Bouabdallaoui:2022wyp,Taghavi:2023ptn}. In particular, the holographic energy density is large during the early phase (due to the small size of the universe), which in turn successfully drives the inflation that also proves to be viable with respect to the recent Planck data. Moreover the holographic realization for an unified scenario from inflation to the dark energy era of the universe in a covariant way has been recently proposed in \cite{Nojiri:2020wmh}. In a different context, particularly in the context of bounce cosmology, holographic energy density proves to be a fruitful agent \cite{Brevik:2019mah} -- actually the presence of holographic energy with a suitable cut-off helps to violate the corresponding null energy condition and allows a bouncing stage of the universe.

Due to such growing cosmological interests of different entropies (like the Tsallis entropy, the R\'{e}nyi entropy, the Barrow entropy, the Sharma-Mittal entropy, the Kaniadakis entropy and the Loop Quantum gravity entropy), the question -- ``does there exist a generalized entropy that can generalize all the known entropies proposed so far?''-- becomes more stronger. In spirit of this, a generalized entropy has been formulated, that proves to reduce all the known entropies for suitable limit of the entropic parameters (see \cite{Nojiri:2022aof,Nojiri:2022dkr,Odintsov:2022qnn,Odintsov:2023qfj,Nojiri:2022nmu,Nojiri:2022ljp}). In particular, two different generalized entropies have been proposed having 6 parameters and 4 parameters respectively, however the conjecture made in \cite{Nojiri:2022dkr} stated that the minimum number of parameters required in a generalized entropy function that can generalize all the aforementioned entropies is equal to four. Both the generalized entropies (with 6-parameters and 4-parameters as well) show a divergent character at the instant of $H = 0$ (where $H$ is the Hubble parameter) which naturally occurs in the context of bounce cosmology at the time of bounce. Such diverging behaviour is common to all the know entropies (like the Tsallis entropy, the R\'{e}nyi entropy, the Barrow entropy, the Sharma-Mittal entropy, the Kaniadakis entropy and the Loop Quantum gravity entropy) as the Bekenstein-Hawking entropy itself diverges at $H = 0$. A possible explanation of this issue is given in \cite{Odintsov:2022qnn,Odintsov:2023qfj} where the authors proposed a singular-free generalized entropy containing 5-parameters, which is singular-free during the entire cosmological evolution of the universe even at $H = 0$ and is able to generalize all the entropies known so far. In this regard, the following conjecture was made -- ``The minimum number of parameters required in a generalized entropy function that can generalize all the known entropies, and at the same time, is also singular-free during the universe’s evolution is equal to five''. Thus as a whole, the generalized entropies having 4-parameters and 5-parameters (constructed in \cite{Nojiri:2022dkr} and in \cite{Odintsov:2022qnn} respectively) are the $minimal~constructions$ of generalized version of entropy depending on whether the universe passes through $H = 0$ during its cosmological evolution.

Having obtained the generalized entropies, the immediate next task is to check their viability, i.e whether the entropic cosmology corresponding to such generalized entropies lead to the correct evolution of universe consistent with the observational data. In the present paper we intend to do this. In particular, we will consider the four parameter generalized entropy (as we are not interested on bounce cosmology in the present work) and will examine its possible implications during the early universe from inflation to reheating stage. Actually after the inflation ends, the universe passes through a non-trivial stage, namely the reheating stage when the inflaton decays to normal particles with a certain decay rate \cite{Martin:2010kz,Dai:2014jja,Cook:2015vqa,Chung:1998rq,Maity:2018dgy,Maity:2018qhi,Haque:2021dha,Ellis:2021kad,Cai:2015soa,DeHaro:2017abf}. Reheating is one of the most important phases of the early universe. It essentially links our standard thermal universe with the pre-thermal state such as the inflationary universe through a complex non-linear process. Over the years, major cosmological observations have given us ample data to understand the theoretical as well as observational aspects of both the thermal and the very early non-thermal inflationary universe cosmology to an unprecedented level. However, our understanding of the intermediate reheating phase is still at a preliminary stage both in terms of both theory and observations. It is generically described by the reheating temperature ($T_\mathrm{re}$) and an effective equation of state (EoS) parameter ($w_\mathrm{eff}$). Until now, both these parameters remain relatively unconstrained, apart from the understanding that the reheating temperature is bounded by the BBN temperature, i.e $T_\mathrm{re} > T_\mathrm{BBN} \sim 10^{-2}\mathrm{GeV}$ (in this regard, one may see \cite{Kawasaki:1999na,Kawasaki:2000en} where the authors suggested a new lower bound on $T_\mathrm{re}$ in a more conservative sense). In the present context, we will study the reheating era from entropic cosmology (or equivalently from holographic realization, as the entropic cosmology can be equivalently mapped to holographic cosmology with suitable holographic cut-offs, see \cite{Nojiri:2021iko,Nojiri:2021jxf}) where the entropic energy density corresponding to the generalized entropy acts as the inflaton which in turn decays to relativistic particles during the reheating phase. The presence of the entropic parameters ensures the continuous evolution of the Hubble parameter from a quasi de-Sitter phase during the inflation to a power law phase during the reheating phase dominated by a constant equation of state (EoS) parameter. Consequently the entropic parameters are constrained coming from the inflation and reheating phenomenology.

The paper is organized as follows: a brief review of generalized entropy and its cosmological field equations are given in Sec.[\ref{sec:2}] and Sec.[\ref{SecI}] respectively. Then the possibility of entropic inflation to entropic reheating and their viability are addressed in Sec.[\ref{sec-inf}] and Sec.[\ref{sec-reheating}] respectively. The case of instantaneous reheating in the present context of generalized entropic cosmology is investigated in Sec.[\ref{sec-reheating}]. The paper ends with some conclusions.

\section{Possible generalizations of known entropies}
\label{sec:2}
In this section, we will briefly discuss the generalized four-parameter entropy function that was initially proposed in \cite{Nojiri:2022dkr}, and which can reproduce the known entropies proposed so far at certain limits of the parameters. In this regard, let us start with the Bekenstein-Hawking entropy \cite{Bekenstein:1973ur,Hawking:1975vcx},
\begin{align}
S = \frac{A}{4G} \,,
\label{BH-entropy}
\end{align}
where $r_\mathrm{H}$ is the horizon radius and $A = 4\pi r_\mathrm{H}^2$ is the area of the horizon.
Consequently, different entropy functions have been introduced
depending on the system under consideration, namely the Tsallis entropy \cite{Tsallis:1987eu}, the R\'{e}nyi entropy \cite{Renyi}, the Barrow entropy \cite{Barrow:2020tzx}, the Sharma-Mittal entropy \cite{SayahianJahromi:2018irq}, the Kaniadakis entropy \cite{Kaniadakis:2005zk} and the Loop Quantum gravity entropy \cite{Majhi:2017zao}.

The growing interest of such different entropies in black hole physics as well as in cosmology naturally raises the question like -- ``Does there exist a unique or generalized entropy that can generalize all these known entropies?'' In spirit of this, some of our authors proposed two different generalized entropy functions containing 6-parameters and 4-parameters respectively \cite{Nojiri:2022aof,Nojiri:2022dkr},
which can generalize all the aforementioned known entropies proposed so far. In particular, the generalized entropies are given by,
\begin{eqnarray}
\mathrm{6~parameter~entropy:}~~~~~\mathcal{S}_\mathrm{6} \left[ \alpha_\pm, \beta_\pm, \gamma_\pm \right]&=&\frac{1}{\alpha_+ + \alpha_-}
\left[ \left( 1 + \frac{\alpha_+}{\beta_+} \, \mathcal{S}^{\gamma_+}
\right)^{\beta_+} - \left( 1 + \frac{\alpha_-}{\beta_-}
\, \mathcal{S}^{\gamma_-} \right)^{-\beta_-} \right] \,,\label{6-entropy}\\
\mathrm{4~parameter~entropy:}~~~~~S_\mathrm{g}\left[\alpha_+,\alpha_-,\beta,\gamma \right]&=&\frac{1}{\gamma}\left[\left(1 + \frac{\alpha_+}{\beta}~S\right)^{\beta} 
 - \left(1 + \frac{\alpha_-}{\beta}~S\right)^{-\beta}\right] \,,
\label{gen-entropy}
\end{eqnarray}
where $S$ is the Bekenstein-Hawking entropy. The entropic parameters of the respective generalized entropy are given in the argument and they are assumed to be positive. It was explicitly demonstrated that both the 4 or 6 -parameter generalised entropy in specific limits give all known entropies proposed so far (i.e, the Tsallis entropy, the R\'{e}nyi entropy, the Barrow entropy, the Sharma-Mittal entropy, the Kaniadakis entropy and the Loop Quantum gravity entropy) \cite{Nojiri:2022aof,Nojiri:2022dkr,Odintsov:2023qfj} . As mentioned in the introduction, that the 4-parameter generalized entropy is the minimal construction for generalized version of entropy, or in other words, the minimum number of parameters required in a generalized entropy function that can generalize all the aforementioned entropies is equal to four. Owing to such minimal construction, we will consider the 4-parameter generalized entropy in the present work. Furthermore, the 4-parameter generalized entropy function in Eq.~(\ref{gen-entropy}) shares the following properties:
(1) $S_\mathrm{g} \rightarrow 0$ for $S \rightarrow 0$. (2) The entropy 
$S_\mathrm{g}\left[ \alpha_+,\alpha_-,\beta,\gamma \right]$ is a monotonically increasing function with $S$. (3) $S_\mathrm{g}\left[ \alpha_+,\alpha_-,\beta,\gamma \right]$ seems to converge to the Bekenstein-Hawking
entropy at certain limit of the parameters given by $\alpha_+ \rightarrow \infty$, $\alpha_- = 0$, $\gamma = \left(\alpha_+/\beta\right)^{\beta}$ and $\beta = 1$.\\

In the context of entropic cosmology, the presence of an entropy function effectively produces an energy density in the Friedmann equations, in effect of which, one may not need to put some extra scalar field by hand to describe the evolution of the universe. This is one of the main advantages of entropic cosmology. In spirit of this and based on the above discussions, here we will consider the four parameter generalized entropy (symbolized by $S_\mathrm{g}$, see Eq.(\ref{gen-entropy})) which is indeed more generalized compared to the 3-parameter entropy function and is also well motivated compared to the six parameter entropy due to less number of parameters, and will examine whether the $S_\mathrm{g}$ can successfully trigger the early phases of the universe, particularly from inflation-to-reheating era.

\section{Cosmology with the 4-parameter generalized entropy} \label{SecI}

We will consider the thermodynamic approach to describe the cosmological behaviour of the universe from the four parameter generalized
entropy function $S_\mathrm{g}$ given in Eq.(\ref{gen-entropy}). For our present purpose, we consider the Friedmann-Lema\^{i}tre-Robertson-Walker (FLRW) space-time with flat spacial section, i.e,
\begin{align}
ds^2=-dt^2+a^2(t)\sum_{i=1,2,3} \left(dx^i\right)^2 \, ,
\label{metric}
\end{align}
where $a(t)$ is known as the scale factor, and correspondingly, $H = \dot{a}/a$ is the Hubble parameter of the universe. The inverse of the Hubble parameter describes the size of the cosmological horizon at a given time. In particular, the radius $r_\mathrm{H}$ of the cosmological horizon is given by $r_\mathrm{H}=\frac{1}{H}$ and the associated Hawking tenparature is defined as $T = \frac{1}{2\pi r_\mathrm{H}} = H/(2\pi)$ \cite{Cai:2005ra}. The entropy contained inside the cosmological horizon having temperature $T$ can be obtained from the Bekenstein-Hawking relation \cite{Padmanabhan:2009vy}. Due to $\dot{H} < 0$ (the overdot represents $\frac{d}{dt}$), the cosmic horizon (and hence the entropy $\propto r_\mathrm{H}^2$) monotonically increases with the time. Such monotonic increasing behaviour of the horizon may be linked with a kind of force applied on the horizon by matter fields inside of the volume $V = \frac{4}{3}\pi r_\mathrm{H}^3$, and generally known as entropic force. Being generated due to the increasing of the horizon radius, the entropic force is expressed by \cite{Komatsu:2013qia,Cai:2010zw}:
\begin{eqnarray}
 F_\mathrm{e} = \rho\frac{dV}{dr_\mathrm{H}}~~,
 \label{new1}
\end{eqnarray}
where $\rho$ symbolizes the energy density of the matter contents. Equivalently, the entropic pressure comes as $p_\mathrm{e} = F_\mathrm{e}/\left(4\pi r_\mathrm{H}^2\right)$ which, for the case of Bekenstein-Hawking entropy, gets proportional to $H^2$ \cite{Komatsu:2013qia}. Thereby the work done by the matter fields, corresponds to the entropic force, can be expressed by,
\begin{eqnarray}
 dW = F_\mathrm{e}dr_\mathrm{H} = \rho dV~~,
 \label{new2}
\end{eqnarray}
owing to which, the first law of thermodynamics turns out to be,
\begin{eqnarray}
 dQ = TdS = -dE + dW~~.
 \label{new3}
\end{eqnarray}
Here $E = \rho V$ is the total energy of the matter contents inside of the cosmological horizon and $T = \frac{H}{2\pi}$ (as given earlier). It may mentioned that the above thermodynamic law can also be directly obtained by using the first Friedmann equation, and the demonstration goes as follows:
\begin{eqnarray}
 S = \frac{\pi}{H^2} = \frac{3}{8\rho} = \frac{3}{8}\left(\frac{V}{E}\right) = S(E,V)~~.
 \label{new4}
\end{eqnarray}
Such expression of entropy in terms of extensive variables $E$ and $V$ immediately leads to its total differential as \cite{Sanchez:2022xfh}:
\begin{eqnarray}
 dS = \frac{2\pi}{H}\left(-dE + \rho dV\right)~~,
 \label{new5}
\end{eqnarray}
which resembles with Eq.(\ref{new3}). Moreover, with the usual Friedmann equations, the thermodynamic law written in Eq.(\ref{new3}) reduces to the unified first law of thermodynamics given by \cite{Akbar:2006kj},
\begin{eqnarray}
 \frac{H}{2\pi}\left(1 + \frac{\dot{H}}{2H^2}\right)dS = -dE + \frac{1}{2}\left(\rho - p\right)dV~~,
 \label{new6}
\end{eqnarray}
where $p$ is the normal pressure of the matter contents. Thus the thermodynamic law of Eq.(\ref{new3}) is consistent with that of in \cite{Akbar:2006kj,Sanchez:2022xfh}.

Eq.(\ref{new3}) clearly demonstrates that the horizon entropy generates due to --- (1) the decreasing of the total internal energy of the matter contents inside of the horizon, represented by the term $-dE$, and (2) the work done of the matter fields on the horizon, given by $\rho dV$. Such decreasing internal energy and the work done may be effectively thought as some energy flux of the matter fields from inside to outside of the horzon, in particular, $-dE + \rho dV = \frac{4}{3}\pi r_\mathrm{H}^3\dot{\rho}dt$ for an infinitesimal time $dt$ (recall that $E = \rho V$). Since the cosmic horizon ($r_\mathrm{H} = H^{-1}$ is a null surface for spatially flat FRW spacetime) divides the observable universe from the unobservable one, the effective energy flux of the matter fields from inside to outside of the horizon can be associated with some information loss which in turn gives rise to entropy of the horizon. Being the work done $\rho dV$ is directly linked with the entropic force; Eq.(\ref{new3}) shows that we already include the possible effects of the entropic force in the thermodynamic law, and consequently, we will use the standard conservation law of the matter fields as $\dot{\rho} + 3H\left(\rho + p\right) = 0$ (in the subsequent calculations) by following the procedure of \cite{Cai:2005ra,Akbar:2006kj,Sanchez:2022xfh}. This is unlike to the scenario of \cite{Komatsu:2013qia} where the possible implications of entropic force has been taken through the conservation law of the matter fields. Here it deserves mentioning that despite the analysis of entropic force, the origin of entropy is a fundamental issue in the entropic gravity scenario (as we also have mentioned in the introduction). In this connection, Sean Carroll et al. in \cite{Carroll:2016lku} argued that it is difficult to find a self-consistent definition of entropy in the framework of thermodynamic gravity. This indicates one missing link in entropic cosmology, and we hope that yet-to-be-built quantum gravity may resolve this issue.

As shown above, the increase of the heat $Q$ in the region comes as \cite{Cai:2005ra}
\begin{align}
\label{Tslls2}
dQ = -\frac{4\pi}{3} r_\mathrm{H}^3 \dot\rho dt = -\frac{4\pi}{3H^3} \dot\rho~dt
= \frac{4\pi}{H^2} \left( \rho + p \right)~dt \, ,
\end{align}
where $\rho$ and $p$ are the matter contents of the universe, which obey the conservation law: $0 = \dot \rho + 3 H \left( \rho + p \right)$.
With the Bekenstein-Hawking entropy, if we apply the first law of thermodynamics $TdS = dQ$ where $T = \frac{H}{2\pi}$ is the Hawking temperature,
one obtains the usual second Friedmann equation as: $\dot H = - 4\pi G \left( \rho + p \right)$. Integrating this immediately leads to
the first Friedmann equation,
\begin{align}
\label{Tslls8}
H^2 = \frac{8\pi G}{3} \rho + \frac{\Lambda}{3} \, ,
\end{align}
where the integration constant $\Lambda$ can be treated as a cosmological constant. 

Instead of the Bekenstein-Hawking entropy of Eq.~(\ref{BH-entropy}), if one uses the generalized entropy in Eq.~(\ref{gen-entropy}), then the first law of thermodynamics gives the following modified Friedmann equation:
\begin{align}
\dot{H}\left(\frac{\partial S_\mathrm{g}}{\partial S}\right) = -4\pi G\left(\rho + p\right) \,.
\label{FRW1-sub}
\end{align}
Clearly with $S_\mathrm{g} \equiv S$, i.e when the Bekenstein-Hawking entropy is used, the above equation gets resemble with the standard Friedmann equation. Due to the form of $S_\mathrm{g}$ in Eq.~(\ref{gen-entropy}), the above equation becomes,
\begin{align}
\frac{1}{\gamma}\left[\alpha_{+}\left(1 + \frac{\pi \alpha_+}{\beta GH^2}\right)^{\beta - 1} 
+ \alpha_-\left(1 + \frac{\pi \alpha_-}{\beta GH^2}\right)^{-\beta-1}\right]\dot{H} = -4\pi G\left(\rho + p\right)~~,
\label{FRW-1}
\end{align}
where we use $S = A/(4G) = \pi/(GH^2)$. By using the conservation relation of the matter contents, i.e., $\dot{\rho} + 3H\left(\rho + p\right) = 0$, we can integrate Eq.(\ref{FRW-1}) to obtain,
\begin{align}
\frac{GH^4\beta}{\pi\gamma}&\,\left[ \frac{1}{\left(2+\beta\right)}\left(\frac{GH^2\beta}{\pi\alpha_-}\right)^{\beta}~
2F_{1}\left(1+\beta, 2+\beta, 3+\beta, -\frac{GH^2\beta}{\pi\alpha_-}\right) \right. \nonumber\\ 
&\, \left. + \frac{1}{\left(2-\beta\right)}\left(\frac{GH^2\beta}{\pi\alpha_+}
\right)^{-\beta}~2F_{1}\left(1-\beta, 2-\beta, 3-\beta, -\frac{GH^2\beta}{\pi\alpha_+}\right) \right] = \frac{8\pi G\rho}{3} + \frac{\Lambda}{3} \,,
\label{FRW-2}
\end{align}
where $2F_1(a,b,c,x)$ denotes the Hypergeometric function and $\Lambda$ is the integration constant (also known as the cosmological constant). Thus as a whole, Eq.~(\ref{FRW-1}) and Eq.~(\ref{FRW-2}) represent the cosmological field equations coming from the four parameter generalized entropy $S_\mathrm{g}$.
With these two pillar equations, we are going to examine the possible implications of $S_\mathrm{g}$ on early evolution of the universe -- this is the subject of next sections.

\section{Inflationary cosmology from the 4-parameter generalized entropy}\label{sec-inf}

Our main aim is to investigate whether the 4-parameter generalized entropy $alone$ can trigger a viable evolution of the universe, and thus we consider that the matter contents and the cosmological constant are absent, i.e $\rho = p = \Lambda = 0$. Moreover during the early stage, the normal matter energy density like the radiation or even the pressure-less dust, if present, are highly diluted due to the inflationary expansion of the universe. However as we will show later that during the reheating process, the entropic energy density corresponding to $S_\mathrm{g}$ gets transformed to the normal matter energy density that are responsible for the standard Big-Bang cosmology after the end of reheating. In effect of $\rho = p = \Lambda = 0$, Eq.~(\ref{FRW-2}) becomes,
\begin{align}
& \left[\frac{1}{\left(2+\beta\right)}\left(\frac{GH^2\beta}{\pi\alpha_-}\right)^{\beta}~2F_{1}
\left(1+\beta, 2+\beta, 3+\beta, -\frac{GH^2\beta}{\pi\alpha_-}\right) \right. \nonumber\\
&\left. \quad +\frac{1}{\left(2-\beta\right)}\left(\frac{GH^2\beta}{\pi\alpha_+}\right)^{-\beta}~2F_{1}
\left(1-\beta, 2-\beta, 3-\beta, -\frac{GH^2\beta}{\pi\alpha_+}\right) \right] = 0 \, .
 \label{FRW-2-inf}
\end{align}
Here we would like to mention that the typical energy scale during inflation is of the order $H \sim 10^{-4}M_\mathrm{Pl}$ (where $M_\mathrm{Pl} \sim 10^{19}\mathrm{GeV}$ is the Planck mass and is related to the Newton's gravitational constant as $M_\mathrm{Pl} = 1/\sqrt{16\pi G}$), due to which, the condition $GH^2 \ll 1$ safely holds at early phase of the universe. In effect, we can use the Taylor series of the Hypergeometric function sitting in Eq.~(\ref{FRW-2-inf}) with respect to the argument containing $GH^2$. As a result, Eq.(\ref{FRW-2-inf}) can be solved, and we obtain a constant Hubble parameter as the solution:
$H = 4\pi M_\mathrm{Pl}\sqrt{\frac{\alpha_+}{\beta}}\left[\frac{(3-\beta)}{(2-\beta)(1-\beta)}\right]$.
For $\frac{\alpha_+}{\beta} \sim 10^{-8}$ and $\beta \lesssim \mathcal{O}(1)$, the above Hubble parameter can be made consistent to the inflationary energy scale, i.e $H \sim 10^{-3}M_\mathrm{Pl}$. The solution of a constant Hubble parameter depicts that the entropic cosmology corresponding to the generalized entropy $S_\mathrm{g}$ results to a de-Sitter inflationary phase during the early universe. However a de-Sitter inflation has no exit, and moreover, the primordial curvature perturbation in a de-Sitter inflation is perfectly scale invariant which is not compatible with the Planck data \cite{Planck:2018jri}. Thus the entropic cosmology corresponding to the $S_\mathrm{g}$ does not lead to a viable inflationary scenario. Here it may be mentioned that getting a constant Hubble parameter as the solution of Friedmann equation(s) is a generic property of entropic cosmology irrespective of entropy function -- this can be easily understood from Eq.(\ref{FRW1-sub}) with $\rho = p = 0$. Owing to this fact and in order to have a quasi de-Sitter inflationary scenario that is also observationally consistent with the Planck data, we consider that the entropic parameters of $S_\mathrm{g}$ are not strictly constant, rather they slowly vary with the cosmic expansion of the universe (i.e, with respect to the cosmic time). The entropic scenario with varying exponent has been studied earlier in \cite{Nojiri:2019skr}, particularly for Tsallis entropy. Such running behavior
of entropic exponent(s) may be argued from quantum gravity, particularly in the case of gravity, if the space-time fluctuates at high energy scales, the degrees of freedom may increase. On the other hand, if gravity becomes a topological theory, the
degrees of freedom may decrease. In particular, here we consider the parameter
$\gamma$ to vary by the following way and the other parameters (i.e., $\alpha_+$, $\alpha_-$ and $\beta$) remain constant with $t$:
\begin{align}
\gamma(N) = \mathrm{exp}\left[\int_{N_\mathrm{f}}^{N}\sigma(N) dN\right]~~.
\label{gamma function}
\end{align}
Here $N$ denotes the e-folding number with $N_\mathrm{f}$ being the total e-folding number of the inflationary era. It may be noted that $\gamma(N_\mathrm{f}) = 1$, i.e $\gamma(N)$ during inflation is continuously connected with that of during reheating by acquiring the value $\gamma(N_\mathrm{f}) = 1$ at the junction from inflation-to-reheating. The function $\sigma(N)$ during inflation takes a different form compared to that of in the reheating era, owing to which, the Hubble parameter continuously evolves from a quasi de-Sitter phase during inflation to a power law phase during the reheating era (dominated by a constant EoS parameter), as we will demonstrate at some stage.

During inflation, $\sigma(N)$ is of the form,
\begin{align}
\sigma(N) = \sigma_0 + \mathrm{e}^{-\left(N_\mathrm{f} - N\right)}~~;~~~~\mathrm{during~inflation} \,,
\label{sigma function}
\end{align}
where $\sigma_0$ is a constant. The term $\mathrm{e}^{-\left(N_\mathrm{f} - N\right)}$ in the above expression becomes effective only around $N = N_\mathrm{f}$, i.e near the end of inflation, and thus the term proves to be useful to make an exit of the inflation under consideration.


In such scenario where $\gamma$ varies with $N$, the Friedmann equation
turns out to be (see the Appendix in Sec.[\ref{sec-appendix}] for detailed calculations),
\begin{align}
 -\left(\frac{2\pi}{G}\right)
\left[\frac{\alpha_+\left(1 + \frac{\alpha_+}{\beta}~S\right)^{\beta-1} + \alpha_-\left(1 + \frac{\alpha_-}{\beta}~S\right)^{-\beta-1}}
{\left(1 + \frac{\alpha_+}{\beta}~S\right)^{\beta} - \left(1 + \frac{\alpha_-}{\beta}~S\right)^{-\beta}}\right]\frac{H'(N)}{H^3} = \sigma(N) \,,
\label{FRW-eq-viable-inf-main1}
\end{align}
on integrating which, we get the Hubble parameter $H = H(N)$ as (see the Appendix in Sec.[\ref{sec-appendix}] for detailed calculations),
\begin{align}
H(N) = 4\pi M_\mathrm{Pl}\sqrt{\frac{\alpha_+}{\beta}}
\left[\frac{2^{1/(2\beta)}\exp{\left[-\frac{1}{2\beta}\int^{N}\sigma(N)dN\right]}}
{\left\{1 + \sqrt{1 + 4\left(\alpha_+/\alpha_-\right)^{\beta}\exp{\left[-2\int^{N}\sigma(N)dN\right]}}\right\}^{1/(2\beta)}}\right] \,.
\label{solution-viable-inf-2}
\end{align}
In the above integration, we use $S = \pi/(GH^2)$, or equivalently, $2HdH = -\frac{\pi}{GS^2}dS$, and recall that $M_\mathrm{Pl} = 1/\sqrt{16\pi G}$ is the four dimensional Planck mass. It is important to note that in Eq.(\ref{FRW-eq-viable-inf-main1}) as well as in Eq.(\ref{solution-viable-inf-2}), we do not use any particular form of $\sigma(N)$ and thus both the equations are valid even after inflation during the reheating era. Here we would like to mention that for $\rho = p = 0$, the first law of thermodynamics results to $dS_\mathrm{g} = dQ/T = 0$ (as $dQ = 0$ in absence of $\rho$ and $p$) or equivalently $S_\mathrm{g} = \mathrm{constant}$. Therefore when the entropic parameters of $S_\mathrm{g}$ (i.e $\alpha_{\pm}$, $\beta$ and $\gamma$) are constant in time, $S_\mathrm{g}$ is the function of only Hubble parameter through the dependence of $S_\mathrm{g} = S_\mathrm{g}(S)$ and thus $dS_\mathrm{g} = 0$ immediately implies the Hubble parameter to be constant (with respect to time) --- as demonstrated in Eq.(\ref{FRW-2-inf}). However if the entropic parameters slowly vary with the cosmic expansion of the universe (as discussed before Eq.(\ref{gamma function})), then $dS_\mathrm{g} = 0$ implies
\begin{eqnarray}
 \left(\frac{\partial S\mathrm{g}}{\partial S}\right)\frac{dS}{dN} + \left(\frac{\partial S\mathrm{g}}{\partial\gamma}\right)\gamma'(N) = 0~~,\label{N1}
\end{eqnarray}
where $S = \pi/\left(GH^2\right)$ is the Bekenstein-Hawking entropy. Since $\frac{dS}{dN}$ is proportional to $H'(N)$, the above equation indicates that the presence of $\gamma'(N)$ effectively acts as an energy density in the modified Friedmann equation corresponding to $S_\mathrm{g}$. Such entropic energy density in turn triggers a non-trivial evolution of the Hubble parameter, as we will demonstrate below.\\

\begin{figure}[!h]
\begin{center}
\centering
\includegraphics[width=3.5in,height=2.5in]{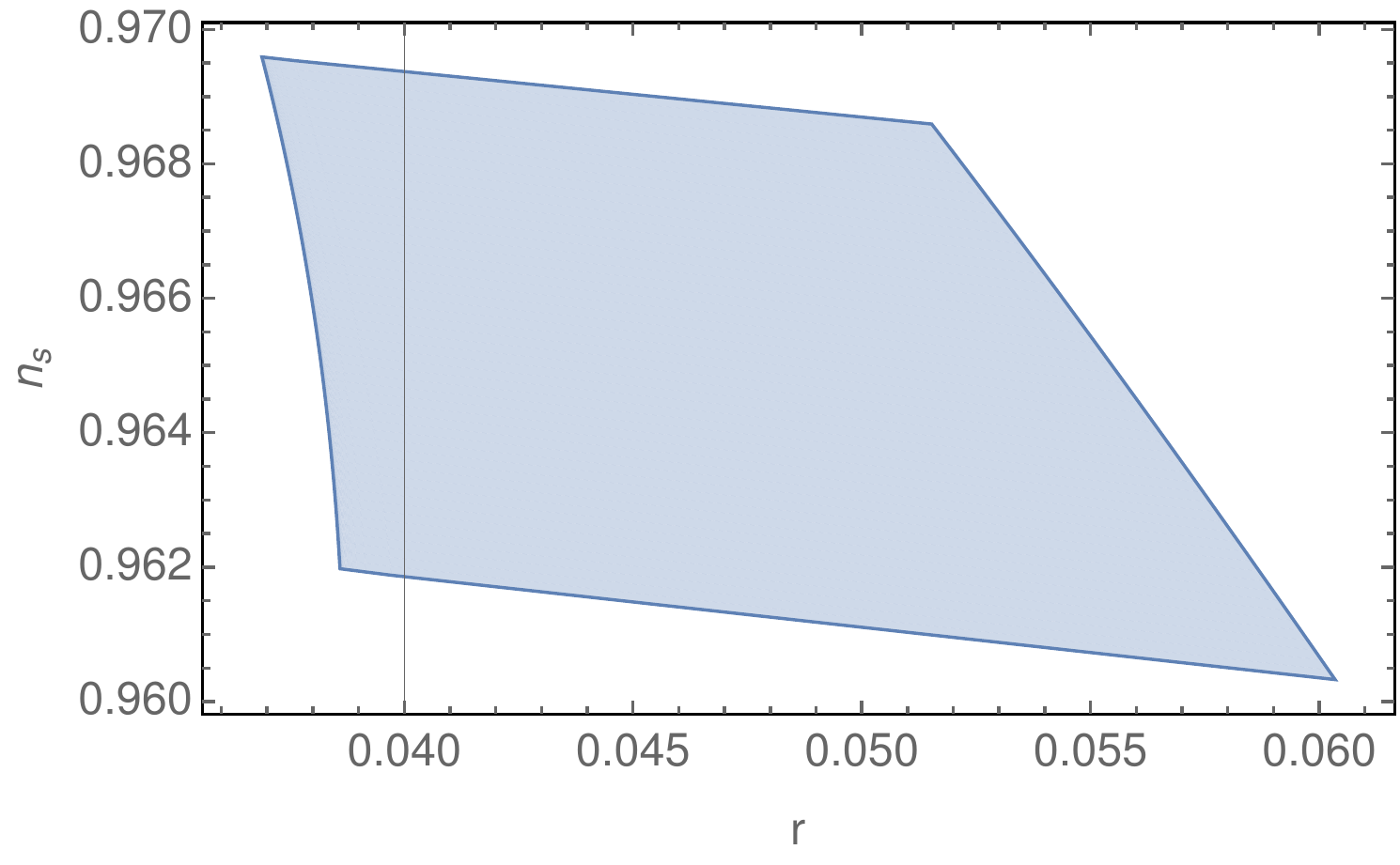}
\caption{Simultaneous compatibility of $n_s$ and $r$ according to Eq.~(\ref{ns final form}) and Eq.~(\ref{r final form}). This corresponds
 to the generalized entropy function $S_\mathrm{g}$ where $\gamma$ varies according to
Eq.~(\ref{gamma function}). In the plot, we take $N_f = 55$, $\sigma_0 = [0.0129,0.0166]$ and $\left(\alpha_+/\alpha_-\right)^{\beta} = [7.5,10^4]$
respectively.}
 \label{plot-observable}
\end{center}
\end{figure}

During inflation, we consider $\sigma(N)$ to be of the form in Eq.(\ref{sigma function}) and consequently we obtain $\int_0^{N}\sigma(N)dN = N\sigma_0 + \mathrm{e}^{-(N_\mathrm{f} - N)} - \mathrm{e}^{-N_\mathrm{f}}$. Eq.~(\ref{solution-viable-inf-2}) represents the solution of the Hubble parameter in the present context of entropic cosmology corresponding to $S_\mathrm{g}$ where $\gamma(N)$ varies
according to Eq.~(\ref{gamma function}) and the other parameters of $S_\mathrm{g}$ are considered to be constant.
Clearly, with a varying $\gamma(N)$, the Hubble parameter is not a constant and thus there is a possibility to have a quasi de-Sitter inflationary era. For this purpose, we now calculate the slow roll parameter defined by $\epsilon = -\frac{\dot{H}}{H^2}$. By using the definition of e-fold number, i.e. $dN = Hdt$, one may write the slow roll parameter in terms of the e-fold number as $\epsilon = -H'(N)/H(N)$ (where the overprime with argument $N$ represents $\frac{d}{dN}$). Due to Eq.(\ref{solution-viable-inf-2}), we determine,
\begin{align}
\epsilon(N) = \frac{\sigma(N)}
{2\beta\sqrt{1 + 4\left(\alpha_+/\alpha_-\right)^{\beta}\exp{\left[-2\int_0^{N}\sigma(N)dN\right]}}} \,.
\label{slow roll parameter}
\end{align}
Consequently we calculate $\epsilon'(N)$ (which is important to determine various observable indices, as we will show later) as,
\begin{align}
\epsilon'(N)/\epsilon(N) = \frac{\sigma(N)}
{1 + \frac{1}{4}\left(\alpha_+/\alpha_-\right)^{-\beta}\exp{\left[2\int_0^{N}\sigma(N)dN\right]}}
+ \frac{\mathrm{e}^{-\left(N_f - N\right)}}{\sigma(N)} \,.
\label{derivative of slow roll parameter}
\end{align}
Due to $\sigma(N) > 0$, Eq.(\ref{derivative of slow roll parameter}) reveals that $\epsilon(N)$ is an increasing function of $N$. Thus the parameters can be fixed in such a way that $\epsilon(N)$ remains less than unity during $N < N_\mathrm{f}$ and reaches to unity at $N = N_\mathrm{f}$. Clearly the condition $\epsilon(N_\mathrm{f}) = 1$ immediately leads to the following relation between the entropic parameters from Eq.(\ref{slow roll parameter}) as follows:
\begin{align}
\beta = \frac{(1 + \sigma_0)}
{2\sqrt{1 + 4\left(\alpha_+/\alpha_-\right)^{\beta}\exp{\left[-2\left(1 + \sigma_0N_f\right)\right]}}} \,,
\label{end of inflation}
\end{align}
where we use $\int_0^{N_f}\sigma(N)dN = 1+\sigma_0N_f$. Before getting involved to scan the parameters, let us evaluate the inflationary observable indices like the spectral tilt for the curvature perturbation ($n_s$) and the tensor-to-scalar ratio ($r$) in the present context. With the expressions of $\epsilon(N)$ and $\epsilon'(N)$ (from Eq.(\ref{slow roll parameter}) and Eq.(\ref{derivative of slow roll parameter})), the final forms of $n_s$ and $r$ are obtained as (see the Appendix in Sec.[\ref{sec-appendix}] for detailed calculations),
\begin{align}
n_s = \bigg[1- 2\epsilon - \frac{2\epsilon'}{\epsilon}\bigg]\bigg|_{h.c} = 1 - \frac{\sigma_0}{\beta\sqrt{1 + \mathrm{exp}\left[2\left(1 + \sigma_0N_\mathrm{f}\right)\right]\left[\left(\frac{1+\sigma_0}{2\beta}\right)^2 - 1\right]}} - \frac{2\sigma_0\left[\left(\frac{1+\sigma_0}{2\beta}\right)^2 - 1\right]}{\mathrm{exp}\left[-2\left(1 + \sigma_0N_\mathrm{f}\right)\right] + \left[\left(\frac{1+\sigma_0}{2\beta}\right)^2 - 1\right]} \,,
\label{ns final form}
\end{align}
and 
\begin{align}
r = 16\epsilon\bigg|_{h.c} = \frac{8\sigma_0}{\beta\sqrt{1 + \mathrm{exp}\left[2\left(1 + \sigma_0N_\mathrm{f}\right)\right]\left[\left(\frac{1+\sigma_0}{2\beta}\right)^2 - 1\right]}}
\label{r final form}
\end{align}
respectively, where the suffix 'h.c' represents the horizon crossing instant of the CMB scale mode ($\sim 0.05\mathrm{Mpc}^{-1}$) on which we are interested. To obtain such expressions of $n_s$ and $r$, we use Eq.(\ref{end of inflation}), i.e the above forms of $n_s$ and $r$ contain the information of $\epsilon(N_\mathrm{f}) = 1$. It is evident that both the observable indices depend on the dimensionless parameters $\sigma_0$ and $\beta$. We can now confront the scalar spectral index and the tensor-to-scalar ratio with the Planck 2018 constraints which gives a bound on the observational indices as: $n_s = 0.9649 \pm 0.0042$ and $r < 0.064$ \cite{Planck:2018jri}. In the present context, the theoretical expectations of $n_s$ and $r$ turn out to be simultaneously compatible with the Planck data for the following ranges of the entropic parameters:
\begin{align}
\sigma_0=&\,[0.0127,0.0166] \,, \quad \left(\alpha_+/\alpha_-\right)^{\beta} \geq 7.5 \,,\nonumber\\
\beta=&\, (0,0.35]\ \mbox{and}\ \left(\alpha_+/\beta\right) \approx 10^{-8} \,,
\label{inf-constraints1}
\end{align}
for $N_\mathrm{f} = 50$. Similarly for $N_\mathrm{f} = 55$ and for $N_\mathrm{f} = 60$, the viable ranges of the parameters are found as follows:
\begin{align}
\sigma_0=&\,[0.0129,0.0166] \,, \quad \left(\alpha_+/\alpha_-\right)^{\beta} \geq 7.5 \,,\nonumber\\
\beta=&\, (0,0.4]\ \mbox{and}\ \left(\alpha_+/\beta\right) \approx 10^{-8} \,,
\label{inf-constraints2}
\end{align}
and
\begin{align}
\sigma_0=&\,[0.013,0.0166] \,, \quad \left(\alpha_+/\alpha_-\right)^{\beta} \geq 7.5 \,,\nonumber\\
\beta=&\, (0,0.4]\ \mbox{and}\ \left(\alpha_+/\beta\right) \approx 10^{-8} \,,
\label{inf-constraints3}
\end{align}
respectively, where the range of $\left(\alpha_+/\alpha_-\right)^{\beta}$ arises due to the inter-relation of Eq.(\ref{end of inflation}). Moreover the consideration of $\frac{\alpha_+}{\beta} \sim 10^{-8}$ leads to the energy scale at the onset of inflation as $H \sim 10^{-4}M_\mathrm{Pl}$. It may be noticed that the viable ranges of the entropic parameters do not change drastically with the total e-fold of the inflationary era. By using Eq.(\ref{ns final form}) and Eq.(\ref{r final form}), below we give the plot of the simultaneous compatibility of $n_s$ and $r$ in respect to the Planck data for $N_\mathrm{f} = 55$, see Fig.[\ref{plot-observable}].

\begin{figure}[!h]
\begin{center}
\centering
\includegraphics[width=3.5in,height=2.5in]{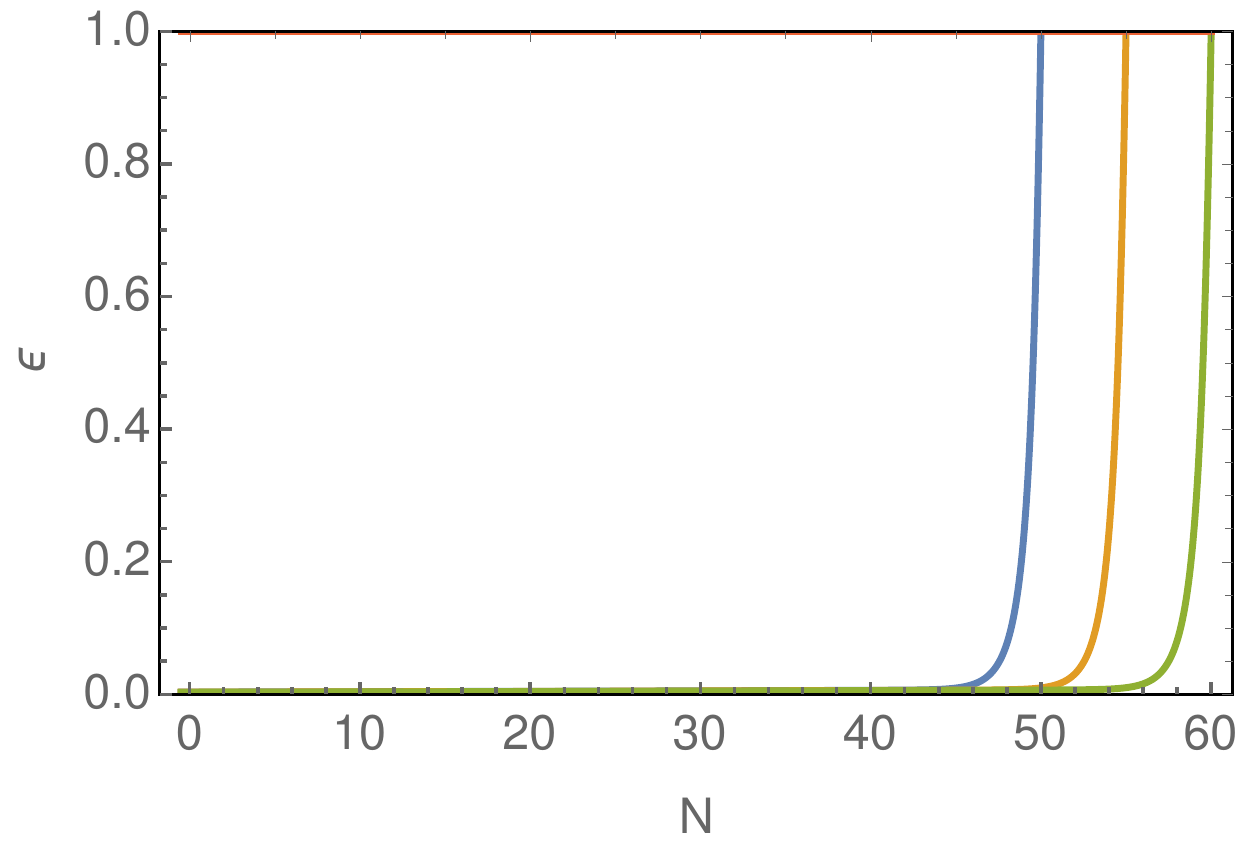}
\caption{$\epsilon(N)$ vs. $N$ from Eq.~(\ref{slow roll parameter}) for three different values of $N_\mathrm{f}$, in particular: $N_\mathrm{f} = 50$ in the blue curve, $N_\mathrm{f} = 55$ in the yellow curve and $N_\mathrm{f} = 60$ in the green curve. Moreover we take $\sigma_0 = 0.015$ and $\left(\alpha_+/\alpha_-\right)^{\beta} = 10$. These three curves reach to unity at their respective $N_\mathrm{f}$.}
\label{plot-sr}
\end{center}
\end{figure}

With a particular set of parametric values from their viable ranges, for instance, let us take $\sigma_0 = 0.015$, and $\left(\alpha_+/\alpha_-\right)^{\beta} = 10$, we give the plot of $\epsilon(N)$ vs. $N$ from Eq.(\ref{slow roll parameter}) for three different values of $N_\mathrm{f} = 50$, $55$ and $60$ respectively; see Fig.[\ref{plot-sr}]. The figure clearly depicts that $\epsilon(N)$ is a monotonic increasing function with $N$ and remains less than unity during $N < N_\mathrm{f}$, and finally reaches to unity at $N = N_\mathrm{f}$. This ensures an inflationary era during the early universe, which gets an exit at a finite e-fold number $N = N_\mathrm{f}$.

Thus as a whole, the entropic cosmology corresponding to the generalized entropy function $S_\mathrm{g}$ where the parameter $\gamma$ is a slowly varying function
of $N$ according to Eq.~(\ref{gamma function}), triggers a viable inflationary scenario during the early universe.
In particular -- (1) the inflation has an exit at a finite e-fold number, which is consistent with the resolution of
horizon and flatness problems. In the above discussions, we present the cases of $N_\mathrm{f} = 50$, $55$ and $60$ respectively. (2) The theoretical expectations of the scalar spectral index for curvature perturbation and the tensor-to-scalar ratio prove to be simultaneously compatible with the recent Planck data for suitable values of the entropic parameters obtained in Eq.~(\ref{inf-constraints1}), Eq.(\ref{inf-constraints2}) and Eq.(\ref{inf-constraints3}) depending on different choices of $N_\mathrm{f}$. At this stage it is important to investigate whether such constraints on the entropic parameters coming from the inflationary requirements support the reheating phase (after the inflation ends) or the parameters get more constrained due to the inputs of the reheating stage -- this is the subject of the next section.

\section{From inflation to reheating}\label{sec-reheating}

After the inflation ends, the universe enters to the reheating era, during which, the entropic energy density corresponding to $S_\mathrm{g}$ decays to relativistic particles with a certain decay width generally considered to be constant and is represented by $\Gamma$. From Eq.(\ref{solution-viable-inf-2}) and Eq.(\ref{FRW-eq-viable-inf-main1}), we immediately write the entropic energy density and the entropic pressure as,
\begin{eqnarray}
 \rho_\mathrm{g} = \frac{3}{16G^2} \left(\frac{\alpha_+}{\beta}\right)
\left[\frac{2\exp{\left[-\int^{N}\sigma(N)dN\right]}}
{1 + \sqrt{1 + 4\left(\alpha_+/\alpha_-\right)^{\beta}\exp{\left[-2\int^{N}\sigma(N)dN\right]}}}\right]^{1/\beta}~~,
 \label{reh-1}
\end{eqnarray}
and
\begin{eqnarray}
 p_\mathrm{g} = -\rho_\mathrm{g} + 16G^2\left(\frac{\rho_\mathrm{g}}{3}\right)^2\sigma(N)
 \left[\frac{\left(1 + \frac{\pi\alpha_+}{\beta GH^2}\right)^{\beta} - \left(1 + \frac{\pi\alpha_-}{\beta GH^2}\right)^{-\beta}}
{\alpha_+\left(1 + \frac{\pi\alpha_+}{\beta GH^2}\right)^{\beta-1} + \alpha_-\left(1 + \frac{\pi\alpha_-}{\beta GH^2}\right)^{-\beta-1}}\right]~~,
 \label{reh-2}
\end{eqnarray}
respectively. Consequently the equation of state (EoS) parameter of entropic energy during the reheating era is given by,
\begin{eqnarray}
 w_\mathrm{g} = \frac{p_\mathrm{g}}{\rho_\mathrm{g}} = -1 + \left(\frac{16G^2}{9}\right)\rho_\mathrm{g}~\sigma(N)
 \left[\frac{\left(1 + \frac{\pi\alpha_+}{\beta GH^2}\right)^{\beta} - \left(1 + \frac{\pi\alpha_-}{\beta GH^2}\right)^{-\beta}}
{\alpha_+\left(1 + \frac{\pi\alpha_+}{\beta GH^2}\right)^{\beta-1} + \alpha_-\left(1 + \frac{\pi\alpha_-}{\beta GH^2}\right)^{-\beta-1}}\right]~~,
 \label{reh-3}
\end{eqnarray}
where $\rho_\mathrm{g}$ is shown in Eq.(\ref{reh-1}). As the entropic energy decays to radiation during the reheating stage, the effective EoS during the same turns out to be,
\begin{eqnarray}
 w_\mathrm{eff} = \frac{3w_\mathrm{g}\rho_\mathrm{g} + \rho_\mathrm{R}}{3\left(\rho_\mathrm{g} + \rho_\mathrm{R}\right)}~~,
 \label{R2-1}
\end{eqnarray}
where $\rho_\mathrm{R}$ represents the radiation energy density. The presence of $\sigma(N)$ in the above expression clearly reveals the possible effects of the generalized entropy during the reheating era. In the present context we consider perturbative reheating caused due to a coupling between entropic energy and relativistic particles, in the same spirit of \cite{Dai:2014jja,Cook:2015vqa}, leading to the decay from entropic energy-to-relativistic particles with a certain decay width ($\Gamma$). It may contain several decay channels to bosons as well as to fermions. After the end of inflation, the Hubble parameter is much larger than the decay width (i.e the condition $H \gg \Gamma$ holds), owing to which, the decaying of entropic energy to radiation is negligible and thus the comoving entropic energy density remains conserved with the cosmological expansion of the universe. However the Hubble parameter continuously decreases and eventually gets comparable to $\Gamma$ when the entropic energy density effectively decays to relativistic particles and indicates the end of reheating. Thus as a whole, the comoving entropic energy density remains conserved (or equivalently, the decaying of the entropic energy is negligible with respect to the Hubble expansion) during $H \gg \Gamma$, and finally, the entropic energy instantaneously decays to relativistic particles at the end of reheating when $H = \Gamma$ satisfies. As a result, the $w_\mathrm{eff}$ from Eq.(\ref{R2-1}) becomes approximately equal to $w_\mathrm{g}$ during the reheating phase, and moreover, $w_\mathrm{eff} = 1/3$ at the end of reheating when $\rho_\mathrm{g} = p_\mathrm{g} = 0$. In particular,
\begin{align}
w_\mathrm{eff}=
\begin{cases}
w_\mathrm{g}~~;~~~\mathrm{during~the ~reheating}~, & \\
1/3~~;~~~\mathrm{at~the~end~of~reheating}~,
\end{cases}
\label{R2-2}
\end{align}
and the end of reheating gets continuously connected to the radiation dominated era. Moreover, in analogy with the standard scalar field cosmology, we consider that the Hubble parameter in the present context follows a power law solution during the reheating era, in particular,
\begin{eqnarray}
 H(t) = m/t~~,
 \label{reh-4}
\end{eqnarray}
during the reheating era, where $m$ is the exponent. Such kind of Hubble parameter during the reheating era also appears in standard cosmology with a scalar field. However in the present work, our motivation is to show that without having any scalar field, the standard cosmology can be achieved purely from thermodynamic analysis based on the four parameter generalized entropy ($S_\mathrm{g}$). Below, we will determine the appropriate form of $\sigma(N)$ corresponding to $H(t) = \frac{m}{t}$ from the governing Eq.(\ref{FRW-eq-viable-inf-main1}) in order to be everything consistent. Due to this expression of $H(t)$, the reheating EoS parameter turns out to be,
\begin{eqnarray}
 w_\mathrm{eff} = -1 - \frac{2\dot{H}}{3H^2} = -1 + \frac{2}{3m}
 \label{reh-5}
\end{eqnarray}
which is actually a constant, resulting to a 'perfect fluid' nature of the entropic energy during the reheating stage. In the above equation, we use the background Friedmann equations as $3H^2 = \rho_\mathrm{g}$ and $2\dot{H} + 3H^2 = -p_\mathrm{g}$, where $\rho_\mathrm{g}$ and $p_\mathrm{g}$ are shown in Eq.(\ref{reh-1}) and Eq.(\ref{reh-2}) respectively. Being $w_\mathrm{eff}$ is a constant, the entropic energy density varies as $\rho_\mathrm{g} \propto a^{-3\left(1 + w_\mathrm{eff}\right)}$ with the scale factor of the universe, or equivalently the comoving entropic energy density, i.e $\rho_\mathrm{g}\times a^{3\left(1 + w_\mathrm{eff}\right)}$, turns out to be conserved.

Here we may recall the standard scalar field cosmology where the scalar field generally oscillates after the end of inflation and triggers the resonance particle production, known as 'preheating' stage \cite{Maity:2018qhi}. In this process the amplitude for the scalar field oscillation gradually decreases, and at a stage, the amplitude becomes considerable low that ceases the preheating. After that, the remaining scalar field energy density perturbatively decays to radiation with a certain decay width --- this is known as 'reheating' \cite{Dai:2014jja,Cook:2015vqa,Maity:2018dgy,Maity:2018qhi}. Here it may be mentioned that unlike to the preheating, the reheating stage does not require an oscillation of the scalar field. In fact, the scalar field oscillation is generally averaged out during the reheating stage, and a result, the scalar field energy behaves as a perfect fluid with constant equation of state during the same. For example --- in the case of $V(\phi) \sim \phi^n$, the scalar field equation of state in the reheating stage turns out to be $\omega_\mathrm{\phi} = \frac{n-2}{n+2}$ which is constant (here $V(\phi)$ is the scalar potential and such potential appears in the alpha attractor model). Moreover the dominance of reheating over the preheating depends on the coupling between the scalar field and radiation, in particular, the reheating becomes the dominant one for a weak coupling of the scalar field (or when $\Gamma \ll H_\mathrm{f}$). Thus for $\Gamma \ll H_\mathrm{f}$, the reheating stage is sufficient to produce adequate radiation in the universe after the BBN.

Coming back to our present model, due to the constant EoS parameter, the entropic energy behaves as a perfect fluid during the reheating stage. Thus the entropic energy does not exhibit any oscillating feature after the end of inflation, and thus the present entropic scenario does not allow a preheating stage. However the entropic energy is allowed to perturbatively decay to radiation, controlled by a certain decay width ($\Gamma$). On analogy of standard scalar field cosmology, let us call this decay process of entropic energy as 'reheating' (or more generally, as 'perturbative reheating'). Similar to \cite{Dai:2014jja,Cook:2015vqa}, the decay from entopic energy to radiation is considered to occur rapidly at the end of the reheating stage when $\Gamma \sim H$ happens. As we will demonstrate below that with an appropriate decay width from entropic energy to radiation, the current model is able to properly reheat the universe without having a preheating stage. Beside the present study, the possibility of a preheating phase and its dominance over the perturbative reheating in entropic cosmology is worthwhile to examine and is expected to address in some future work.\\

\begin{figure}[!h]
\begin{center}
\centering
\includegraphics[width=3.5in,height=2.5in]{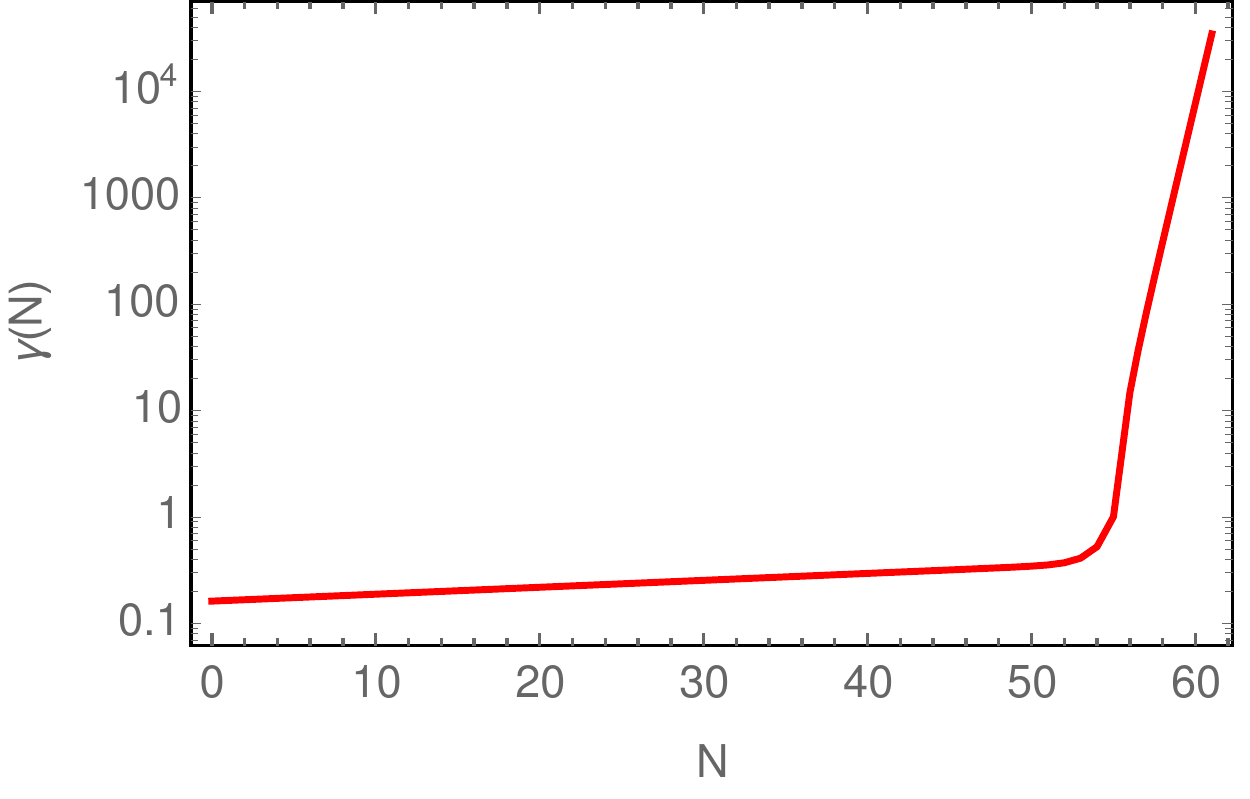}
\caption{$\gamma(N)$ vs. $N$ from inflation to the end of reheating by using Eq.(\ref{gamma function}), Eq.(\ref{sigma function}) and Eq.(\ref{nreh2}). The inflationary duration is taken as $N_\mathrm{f} = 55$, and thus $\gamma(N = 55) = 1$ as expected from Eq.(\ref{gamma function}). Moreover $m = \frac{2}{3}$ in the plot, for which the reheating EoS parameter comes as $w_\mathrm{eff} = 0$. Later in Eq.(\ref{et-14}), we will show that $N_\mathrm{f} = 55$ and $w_\mathrm{eff} = 0$ are consistent provided the entropic parameter $\beta$ lies within $0.20 \lesssim \beta \lesssim 0.40$, and thus we take $\beta = 0.30$ in determining the plot. Consequently the reheating duration for these set of parameter values turns out to be $N_\mathrm{re} \approx 6$, see around Eq.(\ref{et-14}). Therefore the range $0\leq N \leq 55$ represents the inflation and $55 < N \leq 61$ represents the reheating era.}
 \label{plot-gamma}
\end{center}
\end{figure}

Owing to $H(t) = \frac{m}{t}$, the scale factor comes as $a(t) \propto t^{m}$, by which, we may write the Hubble parameter during the reheating era in terms of e-fold number as,
\begin{eqnarray}
 H(N) = H_\mathrm{f}~\mathrm{exp\left[-\left(N - N_\mathrm{f}\right)/m\right]}~~,
 \label{nreh1}
\end{eqnarray}
where recall that $N$ is e-fold number, in particular $N = N_\mathrm{f}$ at the end of inflation and $N = N_\mathrm{f} + N_\mathrm{re}$ when the reheating ends, where $N_\mathrm{f}$ and $N_\mathrm{re}$ represent the durations of inflation and reheating respectively. Having obtained this, we can now reconstruct the required form of $\sigma(N)$ during the reheating stage such that the Hubble parameter follows Eq.(\ref{nreh1}); for this purpose, we use Eq.(\ref{FRW-eq-viable-inf-main1}) (or equally Eq.(\ref{solution-viable-inf-2})) and the reconstructed $\sigma(N)$ is obtained as,
\begin{eqnarray}
 \sigma(N) = \left(\frac{2\pi}{G}\right)\frac{e^{2\left(N - N_\mathrm{f}\right)/m}}{mH_\mathrm{f}^2}
 \left[\frac{\alpha_+\left(1 + \frac{\pi\alpha_+}{\beta GH_\mathrm{f}^2}~e^{2\left(N - N_\mathrm{f}\right)/m}\right)^{\beta-1} + \alpha_-\left(1 + \frac{\pi\alpha_-}{\beta GH_\mathrm{f}^2}~e^{2\left(N - N_\mathrm{f}\right)/m}\right)^{-\beta-1}}
{\left(1 + \frac{\pi\alpha_+}{\beta GH_\mathrm{f}^2}~e^{2\left(N - N_\mathrm{f}\right)/m}\right)^{\beta} - \left(1 + \frac{\pi\alpha_-}{\beta GH_\mathrm{f}^2}~e^{2\left(N - N_\mathrm{f}\right)/m}\right)^{-\beta}}\right]\nonumber\\
 ~~~~~~;~~\mathrm{during~reheating}~~.
 \label{nreh2}
\end{eqnarray}
Thus as a whole, $\gamma(N)$ is of the form of Eq.(\ref{gamma function}) where $\sigma(N)$ during the inflation and during the reheating stages are given by Eq.(\ref{sigma function}) and Eq.(\ref{nreh2}) respectively. Using these, we give the plot of $\gamma(N)$ vs. $N$ from inflation to the end of reheating, see Fig.[\ref{plot-gamma}]. The figure clearly demonstrates that the functional behaviour of $\gamma(N)$ during the reheating changes than that of in the inflationary era, and the change is continuous at the junction of inflation-to-reheating. This in turn ensures the continuous evolution of Hubble parameter from a quasi de-Sitter phase during the inflation to a power law phase during the reheating stage.

\subsection*{Reheating e-folding and reheating temperature}
Let us start by recalling that the Hubble parameter during the reheating era evolves as $H = \frac{m}{t}$ or equivalently $H \propto a^{-\frac{3}{2}\left(1 + w_\mathrm{eff}\right)}$
(where $w_\mathrm{eff} = -1 + \frac{2}{3m}$) in terms of the scale factor of the universe. Thus we may write the Hubble parameter as,
\begin{eqnarray}
 H(a) = H_\mathrm{f}\left(\frac{a}{a_\mathrm{f}}\right)^{-\frac{3}{2}\left(1 + w_\mathrm{eff}\right)}~~,
 \label{et-1}
\end{eqnarray}
where the suffix 'f' with a quantity refers to the quantity at the end of inflation. In this note, we also mention that a suffix 're' denotes the end of reheating. As mentioned earlier that reheating ends when the Hubble parameter gets comparable to the decay width of the entropic energy density, i.e when $H = \Gamma$ satisfies. Thus with the help of Eq.(\ref{et-1}), we may express the decay width in terms of $H_\mathrm{f}$, $w_\mathrm{eff}$ and the e-fold number of the reheating era (defined by $N_\mathrm{re} = \ln{\left(a_\mathrm{re}/a_\mathrm{f}\right)}$) as,
\begin{eqnarray}
 \Gamma = H_\mathrm{f}~\mathrm{exp}\left[-\frac{3}{2}N_\mathrm{re}\left(1 + w_\mathrm{eff}\right)\right]~~.
 \label{et-2}
\end{eqnarray}
Moreover, the Friedmann equation relates $\Gamma$ to the energy density at the end of reheating ($\rho_\mathrm{re}$) by $\Gamma^2 = \frac{1}{3M_\mathrm{Pl}^2}\rho_\mathrm{re}$ when the temperature of the universe is generally symbolized by $T_\mathrm{re}$ (also known as the reheating temperature). Therefore, the reheating e-fold number can be written as,
\begin{eqnarray}
N_\mathrm{re} = \frac{1}{3\left(1 + w_\mathrm{eff}\right)}\ln{\left(\frac{3H_\mathrm{f}^2M_\mathrm{Pl}^2}{\rho_\mathrm{re}}\right)}~~.\label{et-3a}\\
\end{eqnarray}
Owing to the thermalization process, the universe comes to thermal equilibrium at the end of reheating, and thus we have,
\begin{eqnarray}
\rho_\mathrm{re} = \left(\frac{\pi^2g_\mathrm{re}}{30}\right)T_\mathrm{re}^4~~,
 \label{et-3}
\end{eqnarray}
where $g_\mathrm{re}$ denotes the relativistic degrees of freedom. After the reheating ends, the entropy of the universe remains conserved from the end of reheating to the present epoch -- which is due to the fact that the whole energy density corresponding to $S_\mathrm{g}$ gets decay at the end of the reheating. This connects the reheating temperature to the present temperature ($T_0$) of the universe via \cite{Cook:2015vqa},
\begin{eqnarray}
 T_\mathrm{re} = H_\mathrm{i}\left(\frac{43}{11g_\mathrm{re}}\right)^{1/3}\left(\frac{T_0}{k/a_0}\right)e^{-\left(N_\mathrm{f} + N_\mathrm{re}\right)}~~,
 \label{et-4}
\end{eqnarray}
where $\frac{k}{a_0} = 0.05\mathrm{Mpc}^{-1}$ is the CMB scale, $H_\mathrm{i}$ represents the inflationary energy scale (in particular, the Hubble parameter at $N = 0$) and $T_0 = 2.93\mathrm{K}$ symbolizes the present temperature of the universe. Eq.(\ref{et-3}) and Eq.(\ref{et-4}) further give $\rho_\mathrm{re}$ in terms of the inflationary parameters and the reheating e-fold number as,
\begin{eqnarray}
 \rho_\mathrm{re} = H_\mathrm{i}^4\left(\frac{\pi^2g_\mathrm{re}}{30}\right)\left(\frac{43}{11g_\mathrm{re}}\right)^{4/3}\left(\frac{T_0}{k/a_0}\right)^4e^{-4\left(N_\mathrm{f} + N_\mathrm{re}\right)}~~,
 \label{et-5}
\end{eqnarray}
Plugging back the above expression of $\rho_\mathrm{re}$ into Eq.(\ref{et-3a}) along with a little bit of simplification yield the final expression of $N_\mathrm{re}$ as follows:
\begin{eqnarray}
 N_\mathrm{re} = \frac{4}{\left(1 - 3w_\mathrm{eff}\right)}\left\{-\frac{1}{4}\ln{\left(\frac{30}{\pi^2g_\mathrm{re}}\right)} - \frac{1}{3}\ln{\left(\frac{11g_\mathrm{re}}{43}\right)} - \ln{\left(\frac{k/a_0}{T_0}\right)} - \ln{\left[\frac{\left(3H_\mathrm{f}^2M_\mathrm{Pl}^2\right)^{1/4}}{H_\mathrm{i}}\right]} - N_\mathrm{f}\right\}~~.
 \label{et-6}
\end{eqnarray}
Clearly the $N_\mathrm{re}$ is represented in terms of the inflationary parameters (like $N_\mathrm{f}$, $H_\mathrm{I}$) and the reheating EoS parameter ($w_\mathrm{eff}$). With $\frac{k}{a_0} = 0.05\mathrm{Mpc}^{-1} \approx 10^{-40}\mathrm{GeV}$ (where the conversion $1\mathrm{Mpc}^{-1} = 10^{-38}\mathrm{GeV}$ may be useful), $T_0 = 2.93K$ and $g_\mathrm{re} = 100$, one simply gets,
\begin{eqnarray}
 N_\mathrm{re} = \frac{4}{\left(1 - 3w_\mathrm{eff}\right)}\left\{61.6 - \ln{\left[\frac{\left(3H_\mathrm{f}^2M_\mathrm{Pl}^2\right)^{1/4}}{H_\mathrm{i}}\right]} - N_\mathrm{f}\right\}~~.
 \label{et-7}
\end{eqnarray}
The quantity $H_\mathrm{i}$ and $H_\mathrm{f}$, i.e the Hubble parameter at the beginning of inflation and at the end of inflation respectively, can be determined from Eq.(\ref{solution-viable-inf-2}) with $\sigma(N)$ is given by Eq.(\ref{sigma function}). They are obtained as,
\begin{eqnarray}
 H_\mathrm{i} = 4\pi M_\mathrm{Pl}\sqrt{\frac{\alpha_+}{\beta}}
\left[\frac{2}
{\left\{1 + \sqrt{1 + \mathrm{exp}\left[2\left(1 + \sigma_0N_\mathrm{f}\right)\right]\left[\left(\frac{1+\sigma_0}{2\beta}\right)^2 - 1\right]}\right\}}\right]^{1/(2\beta)}~~,
 \label{et-8}
\end{eqnarray}
and
\begin{eqnarray}
 H_\mathrm{f} = 4\pi M_\mathrm{Pl}\sqrt{\frac{\alpha_+}{\beta}}
\left[\frac{4\beta\exp{\left[-\left(1 + \sigma_0N_\mathrm{f}\right)\right]}}
{\left\{1+\sigma_0 + 2\beta\right\}}\right]^{1/(2\beta)}~~,
\label{et-9}
\end{eqnarray}
respectively. To arrive at the above two equations, we use Eq.(\ref{end of inflation}) to replace $\left(\frac{\alpha_+}{\alpha_-}\right)^{\beta}$ with $\beta$, which allow to write $H_\mathrm{i}$ and $H_\mathrm{f}$ in terms $\sigma_0$ and $\beta$.
Plugging back these expressions of $H_\mathrm{i}$ and $H_\mathrm{f}$ into Eq.(\ref{et-7}) and a little bit of simplification yields the final form of reheating e-fold number in the generalized entropic cosmology as follows,
\begin{eqnarray}
 N_\mathrm{re} = \frac{4}{\left(1 - 3w_\mathrm{eff}\right)}\Bigg\{61.6 - \frac{1}{4\beta}\ln{\left[\frac{\beta\exp{\left[-\left(1 + \sigma_0N_\mathrm{f}\right)\right]}\left\{1 + \sqrt{1 + \mathrm{exp}\left[2\left(1 + \sigma_0N_\mathrm{f}\right)\right]\left[\left(\frac{1+\sigma_0}{2\beta}\right)^2 - 1\right]}\right\}^2}
{\left(16\pi^2\alpha_+/3\beta\right)^{\beta}~\left\{1+\sigma_0 + 2\beta\right\}}\right]} - N_\mathrm{f}\Bigg\}~.
 \label{et-10}
\end{eqnarray}
Consequently the reheating temperature can be obtained from Eq.(\ref{et-4}) where $N_\mathrm{re}$ follows the above equation.

Thus as a whole, Eq.(\ref{et-4}) and Eq.(\ref{et-10}) are the pillar equations to examine the reheating constraints in the present context of entropic cosmology. Recall that Eq.(\ref{inf-constraints1}), Eq.(\ref{inf-constraints2}) and Eq.(\ref{inf-constraints3}) depict the inflationary constraints on the entropic parameters for three choices of $N_\mathrm{f} = 50$, $55$ and $60$ respectively. Based on which, we will study the reheating phenomenology for these same choices of $N_\mathrm{f}$ with various values of reheating EoS parameter ($w_\mathrm{eff}$). Due to the reason that inflation ends when the effective EoS parameter of the universe takes the value $= -\frac{1}{3}$, the reheating EoS parameter $w_\mathrm{eff}$, in general, ranges from $-\frac{1}{3}$ to unity. Within this range, we will take $w_\mathrm{eff} = 0, 0.1, 0.2, \frac{2}{3}, 1$ respectively (i.e three values from $w_\mathrm{eff} < 1/3$ and two values from $w_\mathrm{eff} > 1/3$) to study the reheating phenomenology.

\begin{itemize}
 \item \underline{\boldmath{Set-1: $N_\mathrm{f} = 50$}}: For this set, the required plots are shown in Fig.[\ref{plot-R1}] where the left and the right plots depicts $N_\mathrm{re}$ vs. $\beta$ and $T_\mathrm{re}$ vs. $N_\beta$ respectively (by using Eq.(\ref{et-10}) and Eq.(\ref{et-4})). Here we vary $N_\mathrm{re}$ and $T_\mathrm{re}$ with respect to $\beta$ rather than $\sigma_0$, this is because of the fact that the $N_\mathrm{re}$ and $T_\mathrm{re}$ do not change much with respect to $\sigma_0$. Thus in both the plots, we take a fixed value of $\sigma_0 = 0.015$. The left plot of Fig.[\ref{plot-R1}] clearly demonstrates that in order to be $N_\mathrm{re} > 0$, the reheating EoS parameter gets constrained as follows: $-\frac{1}{3} < w_\mathrm{eff} < \frac{1}{3}$ for $0.10 < \beta < 0.35$ and $\frac{1}{3} < w_\mathrm{eff} < 1$ for $0 < \beta < 0.10$ (recall that $\beta = (0, 0.35]$ from inflationary requirements, see Eq.(\ref{inf-constraints1})). On other hand, the right plot of the figure depicts that in order to $T_\mathrm{re} > T_\mathrm{BBN}$, a certain value of $\beta$ further puts constraint on the reheating EoS parameter -- for example, if one takes $\beta = 0.17$ then the reheating EoS parameter must lie in the range $w_\mathrm{eff} < 0.2$ in order to satisfy $T_\mathrm{re} > T_\mathrm{BBN}$. Moreover it is also clear that $\beta$ should be larger than the value $0.05$, otherwise the reheating temperature goes below than the BBN temperature for all possible $w_\mathrm{eff}$. Thus as a whole,
 \begin{eqnarray}
  0&<&\beta < 0.05~~~~~~~\rightarrow~~~~\mathrm{forbidden}~~,\nonumber\\
  0.05&<&\beta < 0.10~~~~~~~\Longrightarrow~~~~\frac{1}{3} < w_\mathrm{eff} < 1~~,\nonumber\\
  0.10&<&\beta < 0.35~~~~~~~\Longrightarrow~~~~-\frac{1}{3} < w_\mathrm{eff} < \frac{1}{3}~~.\label{et-13}
  \end{eqnarray}
  This is broad constraint on $\beta$ from the reheating phenomenology, however as we have just mentioned that a certain value of $\beta$ further restricts $w_\mathrm{eff}$ to obey $T_\mathrm{re} > T_\mathrm{BBN}$.

  \begin{figure}[!h]
\begin{center}
\centering
\includegraphics[width=3.0in,height=2.0in]{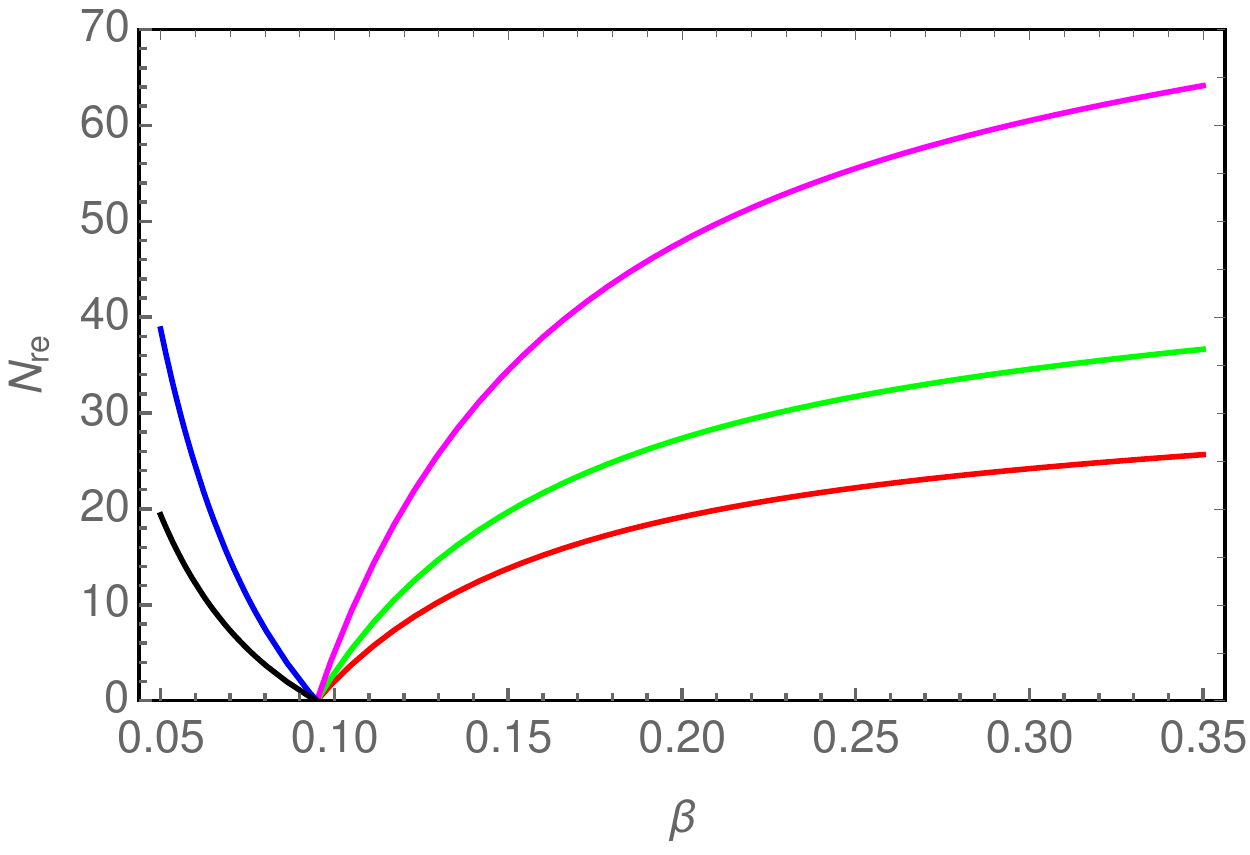}
\includegraphics[width=3.0in,height=2.0in]{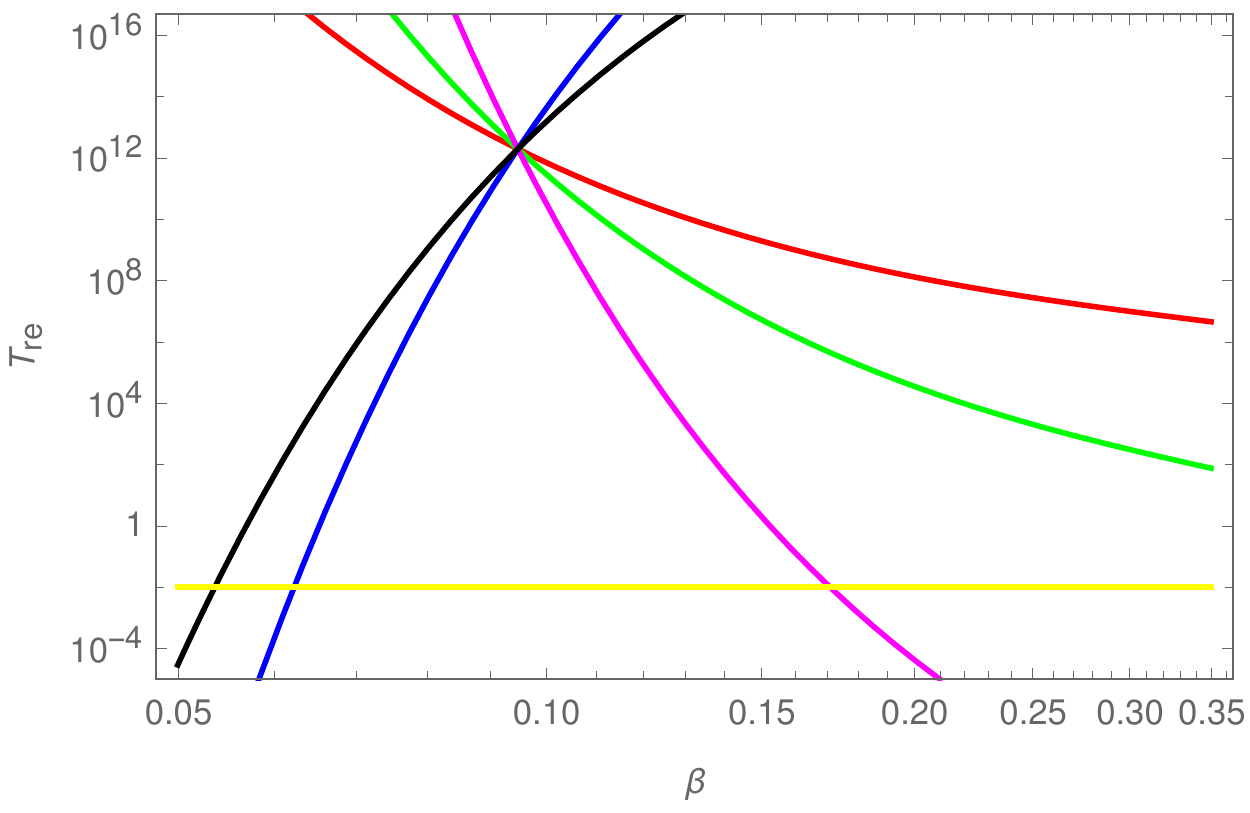}
\caption{{\color{blue}Left Plot}: $N_\mathrm{re}$ vs. $\beta$; {\color{blue}Right Plot}: $T_\mathrm{re}$ vs. $\beta$ for $N_\mathrm{f} = 50$ and various values of $w_\mathrm{eff}$. In both the plots, we consider $\sigma_0 = 0.015$. The reheating EoS parameter is taken as $w_\mathrm{eff} = 0 ~(\mathrm{Red~curve}), 0.1~(\mathrm{Green~curve}), 0.2~(\mathrm{Magenta~curve}), \frac{2}{3}~(\mathrm{Blue~curve}), 1~(\mathrm{Black~curve})$ respectively. Moreover the yellow curve in the right plot is the BBN temperature $\sim 10^{-2}\mathrm{GeV}$.}
\label{plot-R1}
\end{center}
\end{figure}

\end{itemize}

\begin{figure}[!h]
\begin{center}
\centering
\includegraphics[width=3.0in,height=2.0in]{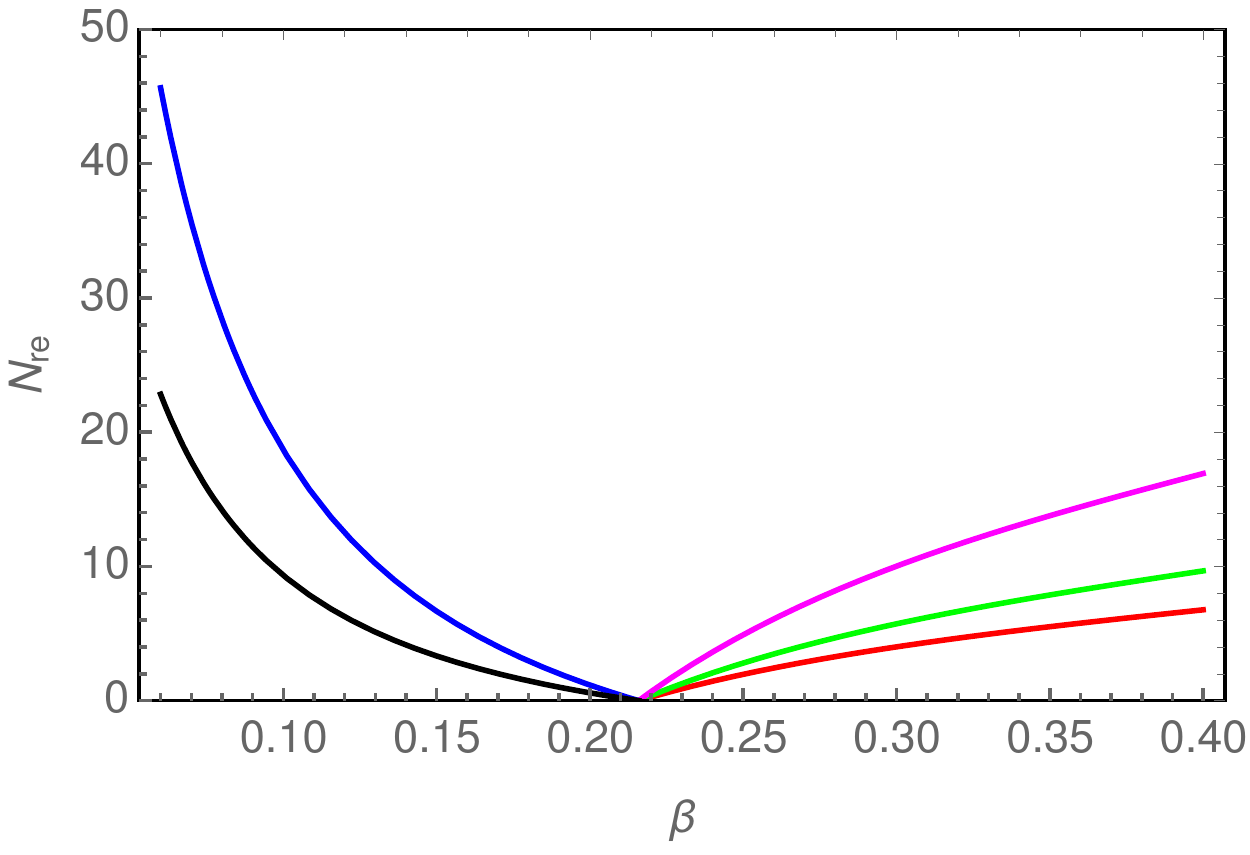}
\includegraphics[width=3.0in,height=2.0in]{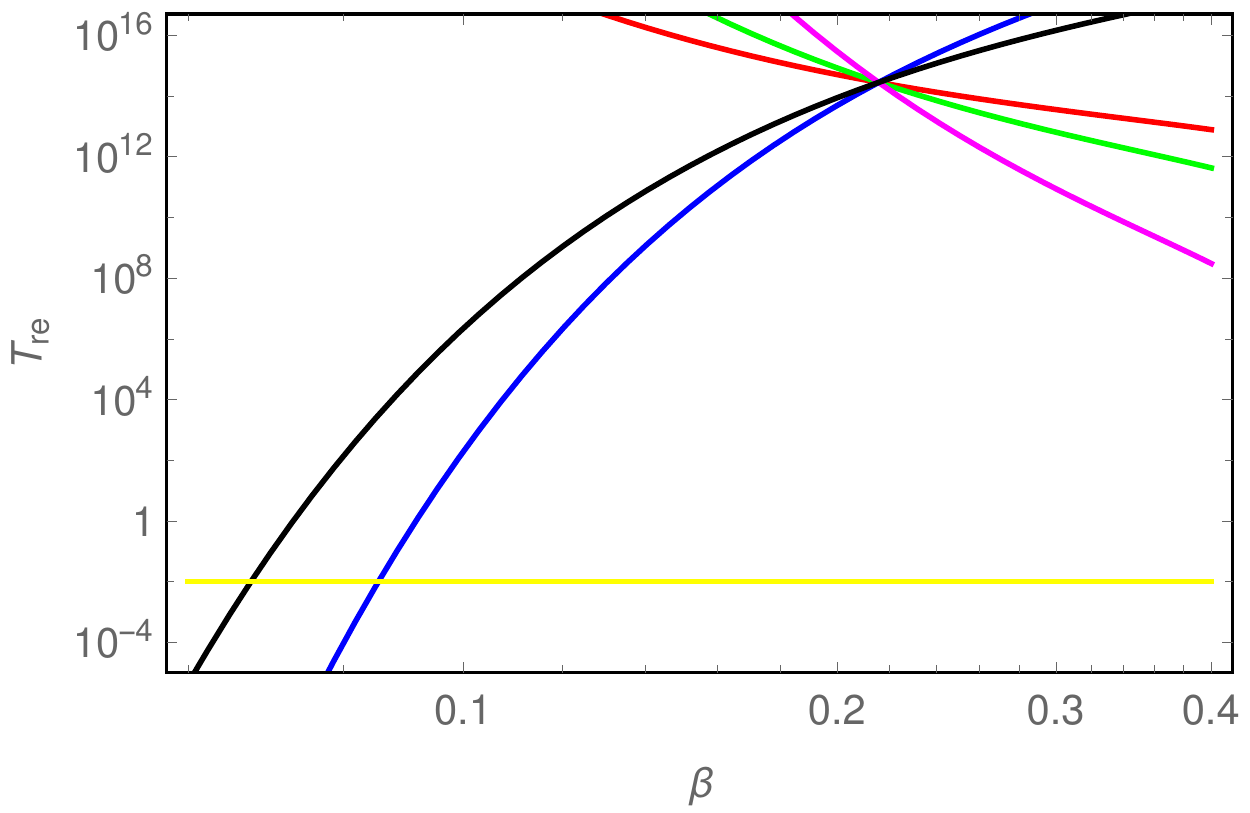}
\caption{{\color{blue}Left Plot}: $N_\mathrm{re}$ vs. $\beta$; {\color{blue}Right Plot}: $T_\mathrm{re}$ vs. $\beta$ for $N_\mathrm{f} = 55$ and various values of $w_\mathrm{eff}$. In both the plots, we consider $\sigma_0 = 0.015$. The reheating EoS parameter is taken as $w_\mathrm{eff} = 0 ~(\mathrm{Red~curve}), 0.1~(\mathrm{Green~curve}), 0.2~(\mathrm{Magenta~curve}), \frac{2}{3}~(\mathrm{Blue~curve}), 1~(\mathrm{Black~curve})$ respectively. Moreover the yellow curve in the right plot is the BBN temperature $\sim 10^{-2}\mathrm{GeV}$.}
\label{plot-R2}
\end{center}
\end{figure}

\begin{itemize}
 \item \underline{\boldmath{Set-2: $N_\mathrm{f} = 55$}}: For this set, we show the required plots in Fig.[\ref{plot-R2}], see $N_\mathrm{re}$ vs. $\beta$ in the left plot and $T_\mathrm{re}$ vs. $\beta$ in the right plot, where we consider a fixed value of $\sigma_0 = 0.015$. From the left plot, it is clear that the reheating EoS parameter gets constrained as follows: $-\frac{1}{3} < w_\mathrm{eff} < \frac{1}{3}$ when $\beta > 0.22$ and $\frac{1}{3} < w_\mathrm{eff} < 1$ when $\beta < 0.22$. Moreover the right plot shows that there is a possibility to be $T_\mathrm{re} < T_\mathrm{BBN}$ for $\beta < 0.06$. Therefore in order to satisfy $T_\mathrm{re} > T_\mathrm{BBN}$, the entropic parameter $\beta$ must lie within $\beta > 0.05$. Thus combining the inflation and reheating phenomenology, what we find is following:
 \begin{eqnarray}
  0&<&\beta < 0.06~~~~~~~\rightarrow~~~~\mathrm{forbidden}~~,\nonumber\\
  0.06&<&\beta < 0.20~~~~~~~\Longrightarrow~~~~\frac{1}{3} < w_\mathrm{eff} < 1~~,\nonumber\\
  0.20&<&\beta < 0.40~~~~~~~\Longrightarrow~~~~-\frac{1}{3} < w_\mathrm{eff} < \frac{1}{3}~~,\label{et-14}
  \end{eqnarray}
 respectively.

\end{itemize}

\begin{itemize}
 \item \underline{\boldmath{Set-3: $N_\mathrm{f} = 60$}}: For this set, we give the required plots in Fig.[\ref{plot-R3}], see $N_\mathrm{re}$ vs. $\beta$ in the left plot and $T_\mathrm{re}$ vs. $\beta$ in the right plot, for $\sigma_0 = 0.015$. The left plot reveals that unlike to the previous two cases, the reheating EoS parameter corresponding to the case $N_\mathrm{f} = 60$ can not be less than $\frac{1}{3}$ for any possible values of $0 < \beta \leq 0.4$, otherwise the reheating e-fold number becomes negative (the inflationary phenomenology requires $0 < \beta \leq 0.4$, see Eq.(\ref{inf-constraints3})). Moreover the requirement of $T_\mathrm{re} > T_\mathrm{BBN}$ further puts a constraint as $\beta > 0.08$ (due to $N_\mathrm{re} < 0$, we do not show the plots of $T_\mathrm{re}$ vs. $\beta$ for $w_\mathrm{eff} < 1/3$). Thus for the present case,
 \begin{eqnarray}
  0&<&\beta < 0.08~~~~~~~\rightarrow~~~~\mathrm{forbidden}~~,\nonumber\\
  0.08&<&\beta < 0.40~~~~~~~\Longrightarrow~~~~\frac{1}{3} < w_\mathrm{eff} < 1~~.\label{et-15}
  \end{eqnarray}
Here it deserves mentioning in the present context, the reheating EoS parameter always larger than $\frac{1}{3}$, i.e $w_\mathrm{eff} > \frac{1}{3}$, for the cases $N_\mathrm{f} \gtrsim 57$.

\begin{figure}[!h]
\begin{center}
\centering
\includegraphics[width=3.0in,height=2.0in]{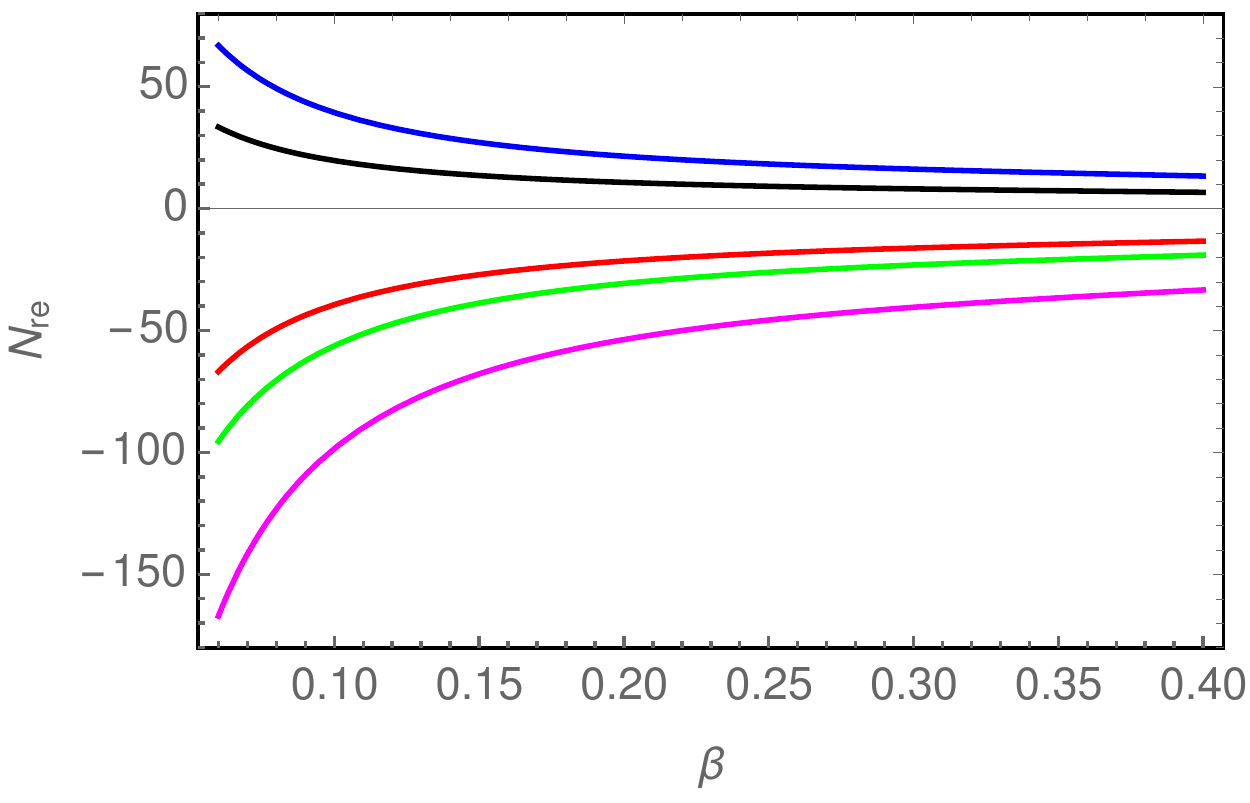}
\includegraphics[width=3.0in,height=2.0in]{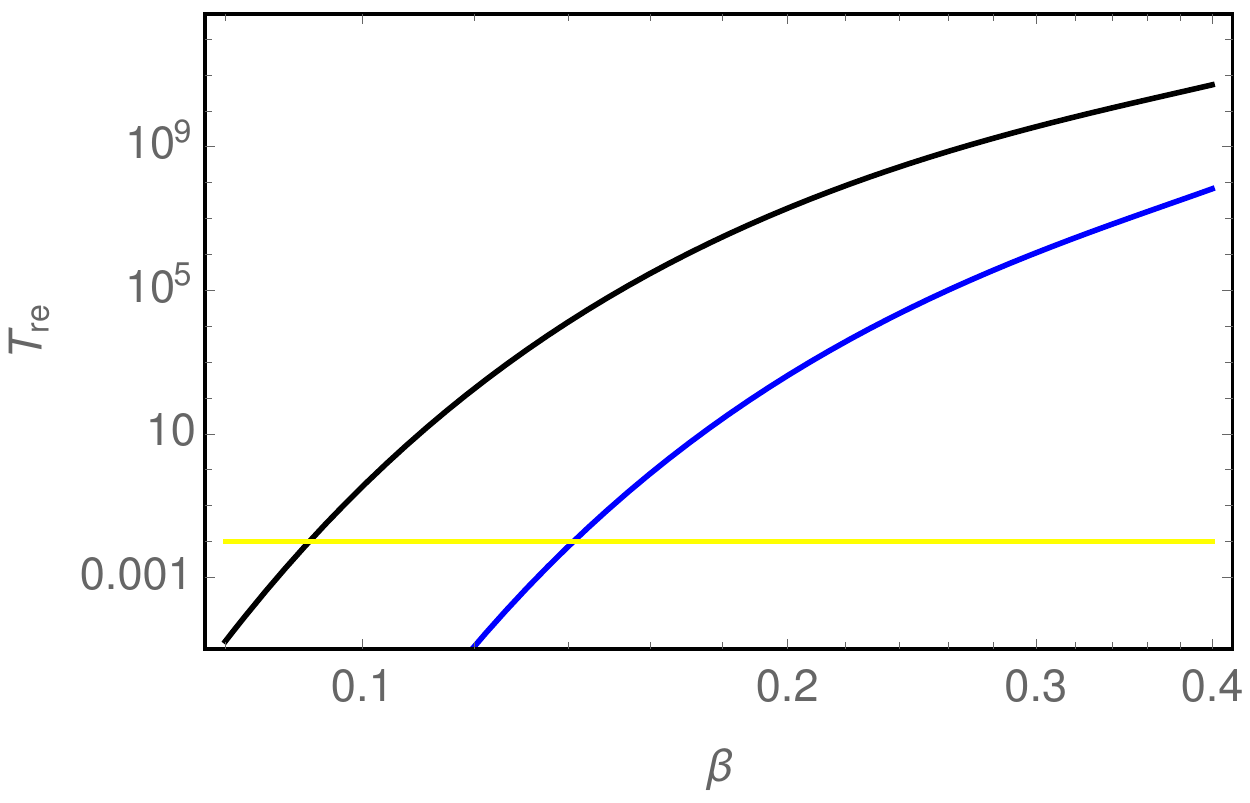}
\caption{{\color{blue}Left Plot}: $N_\mathrm{re}$ vs. $\beta$; {\color{blue}Right Plot}: $T_\mathrm{re}$ vs. $\beta$ for $N_\mathrm{f} = 60$ and various values of $w_\mathrm{eff}$. In both the plots, we consider $\sigma_0 = 0.015$. The reheating EoS parameter is taken as $w_\mathrm{eff} = 0 ~(\mathrm{Red~curve}), 0.1~(\mathrm{Green~curve}), 0.2~(\mathrm{Magenta~curve}), \frac{2}{3}~(\mathrm{Blue~curve}), 1~(\mathrm{Black~curve})$ respectively. Moreover the yellow curve in the right plot is the BBN temperature $\sim 10^{-2}\mathrm{GeV}$.}
\label{plot-R3}
\end{center}
\end{figure}

\end{itemize}

As a whole, we put the viable ranges on the entropic parameters coming from both the inflation and reheating phenomenology in the following Table[\ref{Table-0}].

\begin{table}[h]
  \centering
 {%
  \begin{tabular}{|c|c|c|c|}
   \hline
    Viable choices of $N_\mathrm{f}$ & Viable range of $\beta$ & Viable range of $\left(\frac{\alpha_+}{\alpha_-}\right)^{\beta}$ & Reheating EoS parameter\\

   \hline
  \hline
   (1) Set-1: $N_\mathrm{f} = 50$ & (a) $0.05 < \beta < 0.10$ & $2\times10^{5} < \left(\frac{\alpha_+}{\alpha_-}\right)^{\beta} < 8.5\times10^{5}$ & $\frac{1}{3} < w_\mathrm{eff} < 1$\\
   \hline
     & (b) $0.10 < \beta < 0.35$ & $7.5 < \left(\frac{\alpha_+}{\alpha_-}\right)^{\beta} < 2\times10^{5}$ & $-\frac{1}{3} < w_\mathrm{eff} < \frac{1}{3}$\\
     \hline
    (2) Set-2: $N_\mathrm{f} = 55$ & (a) $0.06 < \beta < 0.22$ & $4\times10^{4} < \left(\frac{\alpha_+}{\alpha_-}\right)^{\beta} < 5\times10^{5}$ & $\frac{1}{3} < w_\mathrm{eff} < 1$\\
   \hline
     & (b) $0.22 < \beta < 0.40$ & $7.5 < \left(\frac{\alpha_+}{\alpha_-}\right)^{\beta} < 4\times10^{4}$ & $-\frac{1}{3} < w_\mathrm{eff} < \frac{1}{3}$\\
     \hline
     (3) Set-3: $N_\mathrm{f} = 60$ & (a) $0.08 < \beta < 0.40$ & $7.5 < \left(\frac{\alpha_+}{\alpha_-}\right)^{\beta} < 3\times10^{5}$ & $\frac{1}{3} < w_\mathrm{eff} < 1$\\
   \hline
     \hline
  \end{tabular}%
 }
  \caption{Viable ranges on entropic parameters coming from both the inflation and reheating phenomenology for three different choices of $N_\mathrm{f}$. In each of the set, the parameter $\sigma_0$ lies within the range $0.013 \lesssim \sigma_0 \lesssim 0.0166$ which arises from the inflationary constraints and does not change from reheating phenomenology.}
  \label{Table-0}
 \end{table}

Thus we may notice that the entropic parameters corresponding to $S_\mathrm{g}$ gets further constrained by the input of the reheating stage. It may be noted that we have considered the lowest bound of reheating temperature to be $\sim 10^{-2}\mathrm{GeV}$. However the authors of \cite{Kawasaki:1999na,Kawasaki:2000en} suggested that $T_\mathrm{re}$ may be lower bounded by $\sim 4\mathrm{MeV}$ in a more conservative sense. If we incorporate this new bound on $T_\mathrm{re}$ in the present context, then the constraints on the entropic parameters get slightly modified.

\subsection*{Possibility of instantaneous reheating}
In the case of instantaneous reheating, the entropic energy density in the present context instantaneously decays to relativistic particles $immediately~after~the~end~of~inflation$. Therefore the reheating e-fold number is zero, i.e $N_\mathrm{re} = 0$, and the reheating temperature from Eq.(\ref{et-4}) comes as,
\begin{eqnarray}
 T_\mathrm{re} = 4\pi M_\mathrm{Pl}\sqrt{\frac{\alpha_+}{\beta}}\left(\frac{43}{11g_\mathrm{re}}\right)^{1/3}\left(\frac{T_0}{k/a_0}\right)e^{-N_\mathrm{f}}
\left[\frac{2}
{\left\{1 + \sqrt{1 + \mathrm{exp}\left[2\left(1 + \sigma_0N_\mathrm{f}\right)\right]\left[\left(\frac{1+\sigma_0}{2\beta}\right)^2 - 1\right]}\right\}}\right]^{1/(2\beta)}~~.
 \label{pos-1}
\end{eqnarray}
Moreover, due to $N_\mathrm{re} = 0$, Eq.(\ref{et-10}) immediately leads to,
\begin{eqnarray}
 61.6 - \frac{1}{4\beta}\ln{\left[\frac{\beta\exp{\left[-\left(1 + \sigma_0N_\mathrm{f}\right)\right]}\left\{1 + \sqrt{1 + \mathrm{exp}\left[2\left(1 + \sigma_0N_\mathrm{f}\right)\right]\left[\left(\frac{1+\sigma_0}{2\beta}\right)^2 - 1\right]}\right\}^2}
{\left(16\pi^2\alpha_+/3\beta\right)^{\beta}~\left\{1+\sigma_0 + 2\beta\right\}}\right]} - N_\mathrm{f} = 0~~,
 \label{pos-2}
\end{eqnarray}
where $H_\mathrm{i}$ and $H_\mathrm{f}$ are obtained in Eq.(\ref{et-8}) and Eq.(\ref{et-9}) respectively.
Therefore the instantaneous reheating is allowed if Eq.(\ref{pos-2}) satisfies for viable $N_\mathrm{f}$. Thus in order to examine the possibility of instantaneous reheating, we need to solve $N_\mathrm{f}$ from Eq.(\ref{pos-2}), and then check whether the solution of $N_\mathrm{f}$ obtained from Eq.(\ref{pos-2}) leads to viable $n_s$ and $r$ with respect to the Planck data. In the following Table [\ref{Table-1}], we examine this by determining $N_\mathrm{f}$ from Eq.(\ref{pos-2}), $(n_s, r)$ from Eq.(\ref{ns final form}) and $T_\mathrm{re}$ from Eq.(\ref{pos-1}).

\begin{table}[h]
  \centering
 {%
  \begin{tabular}{|c|c|c|c|}
   \hline
    Value of $\beta$ & $N_\mathrm{f}$ from Eq.(\ref{pos-2}) & ($n_s$, $r$) from Eq.(\ref{ns final form}) & $T_\mathrm{re}$ (GeV) from Eq.(\ref{pos-1})\\

   \hline
  \hline
   $\beta \approx 0.10$ & $50$ & (0.9648, 0.041) & $5.34\times10^{12}$\\
   \hline
   $\beta \approx 0.22$ & $55$ & (0.9649, 0.042) & $3.16\times10^{14}$\\
   \hline
   \hline
  \end{tabular}%
 }
  \caption{$N_\mathrm{f}$, $(n_s, r)$ and $T_\mathrm{re}$ in the case of instantaneous reheating.}
  \label{Table-1}
 \end{table}

 Table[\ref{Table-1}] clearly demonstrates that the instantaneous reheating in the present context of generalized entropic cosmology is allowed as Eq.(\ref{pos-2}) is satisfied for certain $N_\mathrm{f}$ with viable $(n_s, r)$  and also the reheating temperature is safe from the BBN temperature. In this regard, here it deserves mentioning that the inflationary scenario with $N_\mathrm{f} > 57$ does not support the instantaneous reheating in the current entropic cosmology, as $N_\mathrm{f} > 57$ does  not obey Eq.(\ref{pos-2}) for any value of $\beta$ from its viable range. Thus as a whole, the entropic inflation with $N_\mathrm{f} < 57$ allows instantaneous reheating for suitable $\beta$ -- two such examples are shown in the above table.\\

 Before concluding we would like to mention that this is the first study of reheating in entropic cosmology which indeed proves to be useful in explaining so. The entropic parameters are compatible in describing both the inflation and rehating era of the universe with a smooth evolution of the Hubble parameter at the junction of inflation-to-reheating. Consequently the entropic parameters get tightly constrained (see Table[\ref{Table-0}]), which in turn may provide an unique form (with definite value of the parameters) of the generalized entropy.

\section{Conclusion}

In the context of entropic cosmology, the presence of an entropy function effectively generates an energy density as well as a pressure in the Friedmann equations which are actually obtained from the basic laws of thermodynamics. Based on the generalized entropic construction, here we examine whether the energy density corresponding to the 4-parameter generalized entropy can solely drive the evolution of early universe, particularly from inflation to reheating. The 4-parameter generalized entropy (symbolized by $S_\mathrm{g}$) is the minimal construction of generalized version of entropy that can reduce to all the known entropies proposed so far for suitable limits of the entropic parameters. It turns out that the entropic cosmology corresponding to $S_\mathrm{g}$ successfully triggers an inflationary scenario during the early stage of the universe, and moreover, the inflation has a graceful exit at a finite and adequate amount of e-fold required for resolving the horizon problem. In order to check the viability of the inflation, we calculate the slow roll parameters and consequently evaluate the observable indices like the spectral tilt for curvature perturbation and the tensor-to-scalar ratio at the instant when the CMB scale crosses the horizon. As a result, the theoretical expectations of the observable indices are found to be simultaneously compatible with the recent Planck data for suitable ranges of the entropic parameters. Here it deserves mentioning that the viable ranges of the entropic parameters do not change drastically with the change of total e-fold of the inflationary era within
$N_\mathrm{f} = [50,60]$ considered in the work -- this signifies the robustness of parametric regime in the generalized entropy. After the inflation ends, the entropic energy density corresponding to the $S_\mathrm{g}$ decays to relativistic particles with a certain decay width considered to be a constant (and symbolized by $\Gamma$). The presence of entopic parameter(s) ensure a continuous evolution of the Hubble parameter across the junction at inflation-to-reheating, particularly from a quasi de-Sitter phase during the inflation to a power law phase during the reheating stage. The power law evolution of the Hubble parameter during the reheating phase is due to because the reheating era is dominated by an effective constant EoS parameter that generally lies within $-\frac{1}{3} < w_\mathrm{eff} < 1$. We determine the reheating e-fold number and the reheating temperature, based on which, the reheating phenomenology is examined in the present context of generalized entropic cosmology. Consequently we critically scan the entropic parameters from both the inflation and reheating phenomenology, and it turns out that the viable constraints on the parameters coming from the inflationary phenomenology get more constrained due to the input of the reheating stage. Regarding the reheating EoS parameter in the present entropic cosmology, it can range from $-\frac{1}{3} < w_\mathrm{eff} < 1$ when the inflationary duration is less than 57 e-fold, while the inflation having $N_\mathrm{f} > 57$ seems to be consistent with that reheating stage for which the EoS parameter lies within $\frac{1}{3} < w_\mathrm{eff} < 1$. We further examine the possibility of instantaneous reheating, and as a result, we find that the inflation with $N_\mathrm{f} < 57$ may allow an instantaneous reheating after the end of inflation, in which case, the observable indices simultaneously match with the Planck data and the reheating temperature is also safe from the BBN temperature.

In summary, the generalized entropy proves to be useful in explaining the evolution of early universe from inflation to reheating era. However, our understanding of the fundamental entropy still demands a proper understanding. We hope that the present work of entropic realization (or holographic realization, as the entropic cosmology can be equivalently mapped to holographic cosmology with suitable holographic cut-off) of inflation as well as of reheating era may help in a better understanding of fundamental nature of entropy.

\section{Appendix}\label{sec-appendix}

From the first law of thermodynamics, we have
\begin{eqnarray}
 TdS_\mathrm{g} = -\frac{4}{3}\pi r_\mathrm{H}^3\dot{\rho}dt~~,
 \label{app-1}
\end{eqnarray}
corresponding to the generalized entropy $S_\mathrm{g}$ given in Eq.(\ref{gen-entropy}). Since $\gamma$ varies with $N$, Eq.(\ref{gen-entropy}) indicates that $S_\mathrm{g}$ independently depends on $S$ and $\gamma$ (where $S$ is the Bekenstein-Hawking entropy), and thus
\begin{eqnarray}
 dS_\mathrm{g} = \left\{\left(\frac{\partial S_\mathrm{g}}{\partial S}\right)\frac{dS}{dN} + \left(\frac{\partial S_\mathrm{g}}{\partial\gamma}\right)\frac{d\gamma}{dN}\right\}dN~~.
 \label{app-2}
\end{eqnarray}
Plugging back the above expression of $dS_\mathrm{g}$ into Eq.(\ref{app-1}), along with the fact that $\rho = p = 0$, immediately yields the following equation:
\begin{eqnarray}
 \left(\frac{\partial S_\mathrm{g}}{\partial S}\right)\frac{dS}{dN} + \left(\frac{\partial S_\mathrm{g}}{\partial\gamma}\right)\frac{d\gamma}{dN} = 0~~.
 \label{app-3}
\end{eqnarray}
By using $S = \frac{\pi}{GH^2}$, one gets $\frac{dS}{dN} = -\left(\frac{2\pi}{G}\right)\frac{H'(N)}{H^3}$, and thus Eq.(\ref{app-3}) becomes
\begin{eqnarray}
 -\frac{2\pi}{G}\left(\frac{\partial S_\mathrm{g}}{\partial S}\right)\frac{H'(N)}{H^3} + \left(\frac{\partial S_\mathrm{g}}{\partial\gamma}\right)\gamma'(N) = 0~~.
 \label{app-4}
\end{eqnarray}
Following, we derive various terms present in the above equation:
\begin{itemize}
 \item From Eq.(\ref{gen-entropy}),
 \begin{eqnarray}
  \frac{\partial S_\mathrm{g}}{\partial S} = \frac{1}{\gamma}\left[\alpha_+\left(1 + \frac{\alpha_+}{\beta}~S\right)^{\beta-1}
 + \alpha_-\left(1 + \frac{\alpha_-}{\beta}~S\right)^{-\beta-1}\right]~~,\nonumber
\end{eqnarray}
and
\begin{eqnarray}
 \frac{\partial S_\mathrm{g}}{\partial \gamma} = -\frac{1}{\gamma^2}\left[\left(1 + \frac{\alpha_+}{\beta}~S\right)^{\beta} - \left(1 + \frac{\alpha_-}{\beta}~S\right)^{-\beta}\right]~~.\nonumber
\end{eqnarray}

\item From Eq.(\ref{gamma function}),
\begin{eqnarray}
 \frac{\gamma'(N)}{\gamma(N)} = \sigma(N)~~.\nonumber
\end{eqnarray}

\end{itemize}

With the above expressions, Eq.(\ref{app-4}) immediately leads to,
\begin{eqnarray}
 -\left(\frac{2\pi}{G}\right)
\left[\frac{\alpha_+\left(1 + \frac{\alpha_+}{\beta}~S\right)^{\beta-1} + \alpha_-\left(1 + \frac{\alpha_-}{\beta}~S\right)^{-\beta-1}}
{\left(1 + \frac{\alpha_+}{\beta}~S\right)^{\beta} - \left(1 + \frac{\alpha_-}{\beta}~S\right)^{-\beta}}\right]\frac{H'(N)}{H^3} = \sigma(N) \,,
\label{app-5}
\end{eqnarray}
which is the Eq.(\ref{FRW-eq-viable-inf-main1}) shown in Sec.[\ref{sec-inf}]. Recall that $S = \pi/(GH^2)$, from which, we get $2HdH = -\frac{\pi}{GS^2}dS$.
As a consequence, Eq.~(\ref{app-5}) can be equivalently written as,
\begin{align}
\left[\frac{\alpha_+\left(1 + \frac{\alpha_+}{\beta}~S\right)^{\beta-1} + \alpha_-\left(1 + \frac{\alpha_-}{\beta}~S\right)^{-\beta-1}}
{\left(1 + \frac{\alpha_+}{\beta}~S\right)^{\beta} - \left(1 + \frac{\alpha_-}{\beta}~S\right)^{-\beta}}\right]dS = \sigma(N)dN \,,
\label{app-6}
\end{align}
on integrating which, we obtain the solution of $H = H(N)$ as follows,
\begin{align}
\left(1 + \frac{\pi \alpha_+}{GH^2\beta}\right)^{\beta} - \left(1 + \frac{\pi\alpha_-}{GH^2\beta}\right)^{-\beta}
= \exp{\left[\int_0^{N}\sigma(N)dN\right]}~~.
\label{app-7}
\end{align}
Due to the condition $GH^2 \ll 1$ during the inflationary universe (generally the Hubble parameter during inflation is $\sim 10^{15}\mathrm{GeV}$ and thus $GH^2 \ll 1$), the above equation can be written as,
\begin{eqnarray}
 \left(\frac{\pi \alpha_+}{GH^2\beta}\right)^{\beta} - \left(\frac{\pi\alpha_-}{GH^2\beta}\right)^{-\beta}
= \exp{\left[\int_0^{N}\sigma(N)dN\right]}~~.
\label{app-8}
\end{eqnarray}
Eq.(\ref{app-8}) is an algebraic quadratic equation with respect to $H^{2\beta}$ and thus can be solved for $H = H(N)$ as follows:
\begin{eqnarray}
 H(N) = 4\pi M_\mathrm{Pl}\sqrt{\frac{\alpha_+}{\beta}}
\left[\frac{2^{1/(2\beta)}\exp{\left[-\frac{1}{2\beta}\int^{N}\sigma(N)dN\right]}}
{\left\{1 + \sqrt{1 + 4\left(\alpha_+/\alpha_-\right)^{\beta}\exp{\left[-2\int^{N}\sigma(N)dN\right]}}\right\}^{1/(2\beta)}}\right] \,.
\label{app-9}
\end{eqnarray}
which is the Eq.(\ref{solution-viable-inf-2}). The solution of $H(N)$ helps to determine the slow roll parameter, and consequently, the observable indices like the scalar spectral index and the tensor-to-scalar ratio. The slow roll parameter and its variation with respect to $N$ are determined in Eq.(\ref{slow roll parameter}) and Eq.(\ref{derivative of slow roll parameter}) respectively. These equations immediately lead to the observable indices as,
\begin{align}
 n_s=&\,1 - \frac{\sigma_0}{\beta\sqrt{1 + 4\left(\alpha_+/\alpha_-\right)^{\beta}}} - \frac{8\sigma_0\left(\alpha_+/\alpha_-\right)^{\beta}}
{1 + 4\left(\alpha_+/\alpha_-\right)^{\beta}} \,,\nonumber\\
r=\,&\frac{8\sigma_0}{\beta\sqrt{1 + 4\left(\alpha_+/\alpha_-\right)^{\beta}}}~~,
\label{app-10}
\end{align}
where $\sigma(N = 0) = \sigma_0 + e^{-N_\mathrm{f}} \approx \sigma_0$, and the second term in Eq.(\ref{derivative of slow roll parameter}) (i.e the term containing $e^{-N_\mathrm{f}}$) is neglected due to $e^{-N_\mathrm{f}} \ll 1$. Here we need to recall that $n_s$ and $r$ are determined at the horizon crossing instance of the CMB scale mode ($\sim 0.05\mathrm{Mpc}^{-1}$), which is considered to occur at $N=0$ in the present context. Thus to arrive the above expressions of $n_s$ and $r$, we put $N=0$ in Eq.(\ref{slow roll parameter}) (or in Eq.(\ref{derivative of slow roll parameter})) to get $\epsilon(N)$ and $\epsilon'(N)$ at the horizon crossing. This is the reason why $\sigma(N)$ present in Eq.(\ref{slow roll parameter}) (or in Eq.(\ref{derivative of slow roll parameter})) gets replaced by $\sigma_0$ in deriving the observable indices. On other hand, Eq.(\ref{end of inflation}) gives
\begin{eqnarray}
 \left(\frac{\alpha_+}{\alpha_-}\right)^{\beta} = \frac{1}{4}\left[\left(\frac{1+\sigma_0}{2\beta}\right)^2 - 1\right]\mathrm{exp}\left[2\left(1+\sigma_0N_\mathrm{f}\right)\right]~~,
 \label{app-11}
\end{eqnarray}
by plugging which into Eq.(\ref{app-10}), one can arrive the final forms of $n_s$ and $r$ given in Eq.(\ref{ns final form}) and Eq.(\ref{r final form}) respectively.

\section*{Acknowledgments}

This work was supported in part by MINECO (Spain), project PID2019-104397GB-I00 (SDO). This work was partially supported by the program Unidad de Excelencia Maria
de Maeztu CEX2020-001058-M.

\end{document}